\documentclass[a4paper,11pt]{article}
\pdfoutput=1 

\usepackage{jheppub} 

\usepackage[T1]{fontenc} 
\usepackage[dvipsnames]{xcolor}
\usepackage{dsfont}

\newcommand{\tbf}[1]{\textbf{#1}}
\newcommand{\mds}[1]{\mathds{#1}}

\newcommand*\Pad{ {\bf{P}}_{a}}

\newcommand*\Qid{{\bf{Q}}_{i}}

\newcommand*\Qaid{Q_{a|i}}

\newcommand*{\Qs}{\tbf{Q}}
\newcommand*{\Ps}{\tbf{P}}
\newcommand{\Qst}{\tilde{\Qs}}
\newcommand{\Pst}{\tilde{\Ps}}
\newcommand*{\PP}[2]{\mathds{P}_{#1}^{#2}}
\newcommand*{\QQ}[2]{\mathds{Q}_{#1}^{#2}}
\newcommand*{\qs}{\tbf{q}}
\newcommand*{\ps}{\tbf{p}}
\newcommand*\pad{ {\bf{p}}_{a}}
\newcommand*\qid{{\bf{q}}_{i}}
\newcommand{\qst}{\tilde{\qs}}

\newcommand{\conj}[1]{\bar{#1}}

\newcommand{\eq}[1]{(\ref{#1})}

\newcommand{\br}[1]{\left( #1 \right)}
\newcommand{\be}[1]{\left[ #1 \right]}
\newcommand{\cD}{\bar{D}}
\newcommand{\norm}[1]{\left| #1 \right|}

\newcommand*{\PO}{\text{P}}
\newcommand*{\PhiDotn}{\vec{\Phi}\cdot\vec{n}}

\usepackage{amsmath}
\DeclareMathOperator{\tr}{tr}

\usepackage{graphicx}

\title{\boldmath Excited States of One-Dimensional Defect CFTs from the Quantum Spectral Curve}

\author{David Grabner,$^\phi$}
\author{Nikolay Gromov,$^{\phi,\theta}$}
\author{Julius Julius$^\phi$}

\affiliation{$^\phi$Department of Mathematics, King's College London,
The Strand, London WC2R 2LS, UK}
\affiliation{$^\theta$St.Petersburg INP, Gatchina, 188 300, St.Petersburg,
  Russia}

\emailAdd{david.grabner$\bullet$kcl.ac.uk, nikolay.gromov$\bullet$kcl.ac.uk, julius.julius$\bullet$kcl.ac.uk}

\abstract{
We study the anomalous dimension of the cusped Maldacena-Wilson line in planar $\mathcal{N} = 4$ Yang-Mills theory with scalar insertions using the Quantum Spectral Curve (QSC) method.
In the  straight line limit
we interpret the excited states of the QSC as insertions of scalar operators coupled to the line. Such insertions were recently intensively studied in the context of the 
one-dimensional defect CFT.
We compute a five-loop perturbative result analytically at weak coupling and the first four orders in the $1/\sqrt\lambda$ expansion at strong coupling,
confirming all previous analytic results. In addition, we find the non-perturbative spectrum numerically and show that it interpolates smoothly between the weak and strong coupling predictions. 
}

\begin{document} 
\maketitle
\flushbottom

\section{Introduction}
\label{sec:intro}

The AdS$_5$/CFT$_4$ correspondence \cite{Maldacena:1997re, Witten:1998qj, Gubser:1998bc} conjectures a duality between four-dimensional $\mathcal{N}=4$ supersymmetric Yang-Mills (SYM) theory and type IIB superstring theory on $AdS_{5}\times S^{5}$. The correspondence is of the strong-weak type, so that the strong coupling regime of one theory is mapped to the weak coupling regime of the other, and vice versa. To test this duality, one has to calculate a physical quantity on both sides, finding agreement. While this is not usually possible using perturbative methods, due to the strong-weak nature of the correspondence, integrability \cite{Beisert:2010jr,Dorey:2019gkd} provides non-perturbative tools for exactly this, allowing one to test the AdS/CFT correspondence in highly non-trivial ways. 

Integrability in gauge theories was first discovered in QCD in~\cite{Lipatov:1993yb, Faddeev:1994zg}. Later, it was found in a different context in $\mathcal{N} = 4$ SYM in the seminal paper \cite{Minahan:2002ve}. The development of integrability techniques led to the discovery of the Quantum Spectral Curve (QSC) 
\cite{Gromov:2013pga, Gromov:2014caa}, describing the spectrum of local operators (see \cite{Gromov:2017blm,Kazakov:2018hrh,Levkovich-Maslyuk:2019awk} for reviews). 
In \cite{Gromov:2015wca}, a numerical method was proposed to solve the QSC. This algorithm enables one to probe 
the finite coupling regime, thus allowing the QSC to reach its truly non-perturbative potential.
The main objects of interest are the so-called Q-functions. 
They are obtained as a solution of a Riemann-Hilbert type problem, determined by asymptotic information, analyticity constraints, functional relations, and certain linear monodromy equations across a branch cut, called gluing conditions.  

In this paper we consider the Maldacena-Wilson line~\cite{Maldacena:1998im,Erickson:2000af}. For special shapes like a straight line, the Wilson loop is $1/2$-BPS and the expectation value can be computed exactly by resumming the relevant Feynman diagrams \cite{Erickson:2000af, Drukker:2000rr} or by using methods like localisation\footnote{Strictly speaking one has to first map the line to a circle by a conformal transformation for those results to apply.} \cite{Pestun:2007rz}.
This setup can be generalised by introducing a cusp into the Wilson line, which results in divergences. 
These divergences are controlled by the so-called cusp anomalous dimension, which admits a perturbative expansion at both weak and strong coupling \cite{Makeenko:2006ds, Drukker:2011za, Correa:2012nk, Henn:2013wfa}. 
Cusps in Wilson lines behave very much like local single-trace operators, and a set of boundary Thermodynamic Bethe Ansatz (TBA) equations was found for the cusp anomalous dimension in \cite{Correa:2012hh, Drukker:2012de}.
In the near-BPS limit, when the line becomes almost straight, these equations can be solved analytically \cite{Correa:2012at,Gromov:2012eu,Gromov:2013qga}. However, in general the TBA equations have a number of technical problems, which do not allow one to use them efficiently.
The QSC method was applied to compute the cusp anomalous dimension non-perturbatively in \cite{Gromov:2015dfa}, where the classical strong coupling result \cite{Drukker:2011za, Gromov:2012eu} was reproduced numerically. 

Another way to modify the 1/2-BPS Wilson line is to insert operators along its contour \cite{Drukker:2006xg}. Such an infinite straight Wilson line with insertions can be considered as a superconformal defect with a one-dimensional CFT living on it \cite{Giombi:2017cqn}. The expectation value of operator insertions in a $1/2$-BPS Wilson line in the four-dimensional theory thus corresponds to correlation functions in the one-dimensional  CFT.
One-dimensional CFTs are of interest for the boostrap programme due to their simplicity~ \cite{Mazac:2018mdx, Mazac:2018ycv,Dolan:2011dv,Mazac:2016qev}.
Furthermore, this provides an ideal playground to investigate the AdS$_2$/CFT$_{1}$ correspondence, and has been intensively studied recently \cite{Giombi:2017cqn,Beccaria:2017rbe,Kim:2017sju,Cooke:2017qgm}. 

In this paper we discuss both the cusped Wilson line and the defect CFT living on the infinite straight Wilson line using the QSC. This allows us to develop an integrability description for the spectrum of the one-dimensional CFT. 

\paragraph{Setup and notations.} Consider two Wilson lines in $\mathcal{N} = 4$ SYM, intersecting at an arbitrary angle $\phi$. 
The coupling to the scalars of the theory along the lines is parametrised by two unit vectors, $\vec{n}$ and $\vec{n}_{\theta}$, such that $\vec n \cdot \vec{n}_{\theta}=\cos\theta$.
\begin{figure}[ht]
    \centering
    \includegraphics[scale=0.75]{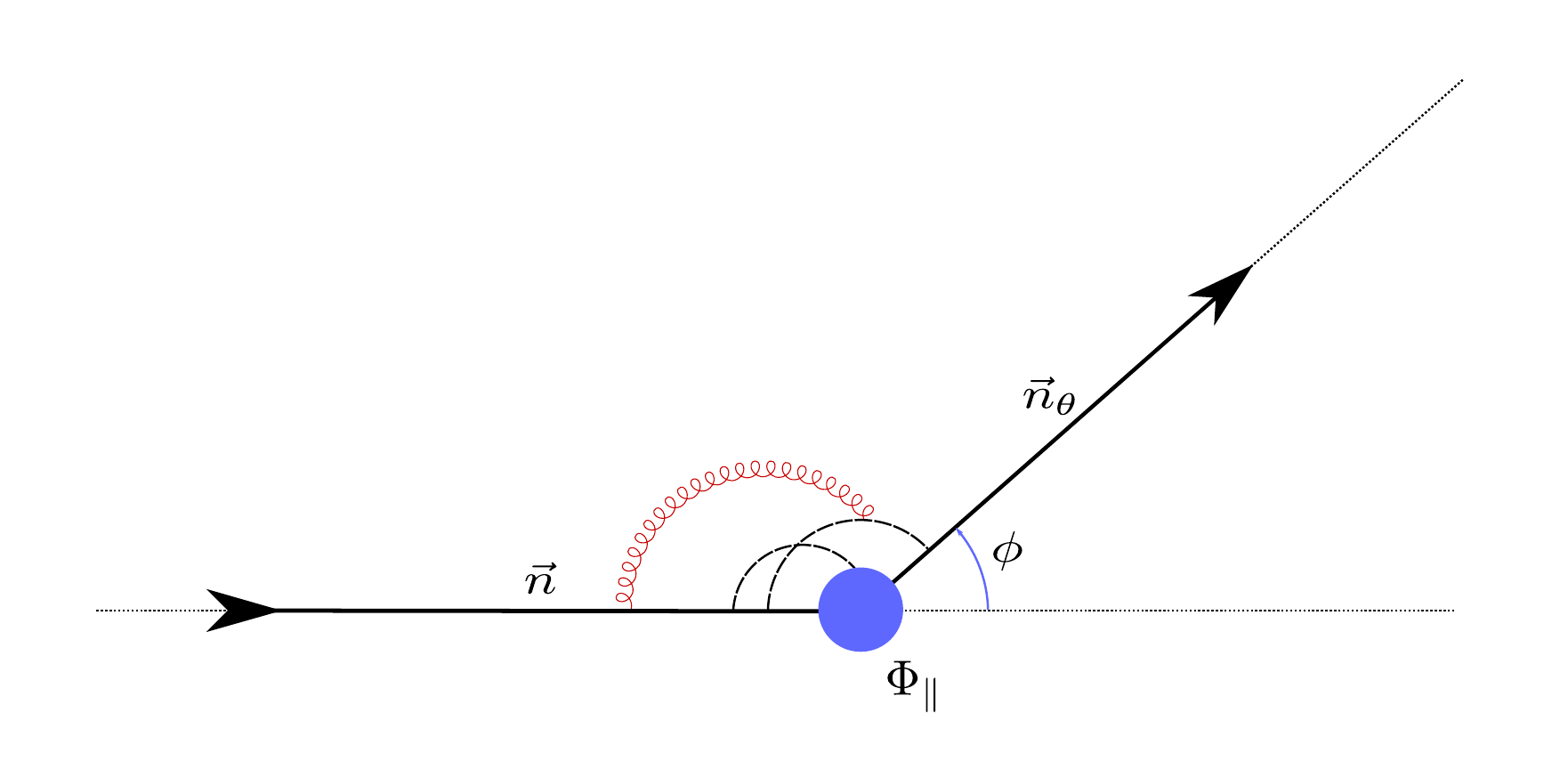}
    \caption{\small The excited cusp: A Feynman diagram for the insertion of $\Phi_\parallel$ (denoted by the blue dot) at the cusp of two semi-infinite Wilson rays, forming a cusp with angle $\phi$. The angle between the scalars coupled to the two rays is $\theta$. Dashed lines represent scalar propagators, whereas gluon propagators are depicted by the usual spiral. In contrast to the insertions considered in \cite{Drukker:2006xg,Drukker:2012de,Correa:2012hh}, the scalar insertion in this case can interact with the scalars that couple to the Wilson line. Therefore, in addition to having the gluon-scalar-scalar vertex, we are also allowed to have a scalar propagator that goes from the insertion to the line.}
    \label{fig:cusp}
\end{figure}
Explicitly, we have
\begin{align}\label{eqn:WL}
W = \tr \PO\exp\left(\int_{-\infty}^{0}dt(i A\cdot \dot{x} + \vec{\Phi}\cdot\vec{n}|\dot{x}|)\right)\times \PO\exp\left(\int_{0}^{\infty}dt(i A\cdot \dot{x}_{\phi}+\vec{\Phi}\cdot\vec{n_{\theta}}|\dot{x}_{\phi}|)\right)\ ,
\end{align}
where $\vec{\Phi}$ is a vector made out of the six scalars, and $x(t)$ and $x_{\phi}(t)$ are straight lines that form an angle $\phi$ at the cusp. This setup is depicted in figure \ref{fig:cusp}. The expectation value of this observable diverges as
\begin{align}\label{eqn:cuspdef}
    \langle W \rangle \sim \left(\frac{\Lambda_\text{IR}}{\Lambda_\text{UV}}\right)^{\Delta}, 
\end{align}
where $\Lambda_\text{IR/UV}$ are infrared and ultraviolet cutoffs, respectively, and $\Delta$ is called the cusp anomalous dimension. 
 
Now consider the expectation value of a cusped Wilson line with $L$ scalar fields inserted at the cusp. Let us decompose the six-dimensional space of scalars as
\begin{equation}
    \mathbb{R}^6=\mathbb{R}^4_\perp \oplus \mathbb{R}^2_\parallel\, , \quad\quad \mathbb{R}^2_\parallel=\text{span}(n,n_\theta)\, ,
\end{equation}
where the corresponding insertions will be called orthogonal and parallel, respectively. First, consider orthogonal insertions. Here, a number $L$ of scalars is inserted at the cusp, denoted by  $\Phi_\perp = \vec{\Phi}\cdot \vec{n_\perp}$, such that $\vec{n}_\perp$ is orthogonal to both $\vec{n}$ and $\vec{n}_{\theta}$. The cusp anomalous dimension is obtained by considering the case without insertions, i.e.~$L = 0$. While the QSC was originally formulated for local single-trace operators of $\mathcal{N}=4$ SYM, it was adapted in \cite{Gromov:2015dfa} to the cusped Wilson line with $L$ orthogonal insertions.

In addition to the orthogonal scalar insertions considered above, it is possible to insert a combination of scalars that couple to the two Wilson lines. In this paper we study such parallel scalar insertions in the cusped Wilson line, which we denote as $\Phi_\parallel$. Due to operator mixing the spectrum is found by diagonalising the mixing matrix. However, the explicit form of the eigenstates is not known in general. While they contain a combination of the scalars $\Phi\cdot n$ and $\Phi\cdot n_\theta$, there could also be other fields present. The explicit form of the insertions was derived in the ladders limit in \cite{Cavaglia:2018lxi}, where only a special class of Feynman diagrams needs to be considered, as originally observed in \cite{Erickson:2000af}.
We review this limit in section \ref{sec:lad}, and extend the QSC description of the first excited state to the case of general angles. In the case of the straight line limit, with both $\phi\to 0$ and $\theta \to 0$, there is a number of perturbative results available \cite{Alday:2007he, Bruser:2018jnc, Giombi:2017cqn} studying the single insertion of $\vec{\Phi}\cdot \vec{n}$, which we refer to as the first excited state. This observable is the main focus of this paper.\footnote{The case with finite $\phi$ and $\theta$
is also of interest from the one-dimensional defect CFT perspective. It can be interpreted as a {\it colour-twist} operator described in \cite{Cavaglia:2020hdb} (see also \cite{Dorn:2020meb}).}

These types of insertions were hitherto outside the scope of the integrability framework of~\cite{Correa:2012hh, Drukker:2012de}. 
However, it was observed in \cite{Cavaglia:2018lxi} that such observables can be naturally studied using the QSC. 
For the case of non-zero angles it was noticed that these insertions satisfy the same QSC equations as the cusped state without insertions, but differ in their classical scaling dimension $\Delta_0=L$. In this paper we study the case of a single insertion of $\Phi_\parallel$, and develop its QSC description in the presence of a cusp outside the ladders limit, i.e.~for generic values of the angles $\phi,\, \theta$.

\begin{figure}[t]
    \centering
    \includegraphics[scale=1]{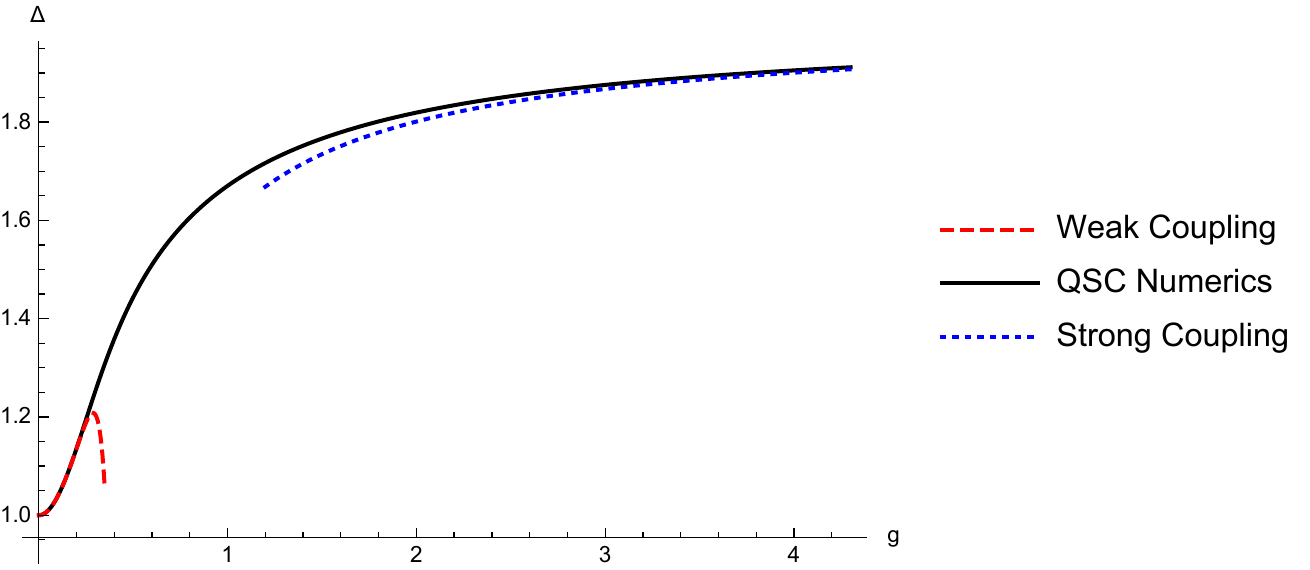}
    \caption{Numerical data for the case with $\phi=\theta=0$, depicted by the solid black line. The dashed red line depicts our five-loop result at weak coupling, see eq.~\eqref{eqn:upto5loops}. Our data reproduces the strong coupling result derived in \cite{Giombi:2017cqn} depicted by the dotted blue line. At finite coupling it interpolates between them.}
    \label{fig:num}
\end{figure}

\paragraph{Main results.} After finding the correct solution of the QSC for generic angles $\phi,\, \theta$, we take both of them to zero. This is equivalent to computing the two-point correlator of this operator with its conjugate in the corresponding defect CFT. In the QSC picture, this amounts to ``untwisting'' the Q-functions, a process which changes their asymptotics in a non-trivial way. We compute the anomalous dimension $\Delta$ of this insertion
at five loop orders
$\Delta\simeq 1+g^2\gamma_1+\dots+g^{10} \gamma_5$,\footnote{$g$ is related to the 't Hooft coupling $\lambda$ as $g=\frac{\sqrt{\lambda}}{4 \pi}$.}
with $\gamma_1=4\ ,\gamma_2=-16$  known previously
\cite{Alday:2007he,Bruser:2018jnc},
and our new result given by
 \begin{equation}\label{eqn:upto5loops}
\begin{split}
    \gamma_3&=-\frac{56 \pi ^4}{45}+128\ ,\\
    \gamma_4&=\frac{272}{135}\pi^6+\frac{128}{3}\pi^2-\frac{64}{3}\pi^2 \zeta_3 +128 \zeta_3 -160 \zeta_5-1280\ , \\
    \gamma_5 &= -\frac{7328}{2835}\pi^8 -\frac{64}{2835}\pi^6 -\frac{896}{45}\pi^4 -\frac{2560}{3}\pi^2 +\frac{64}{3}\pi^4 \zeta_3 +\frac{512}{3}\pi^2 \zeta_3 +\frac{448}{3}\pi^2 \zeta_5 \\
    &- 384 (\zeta_3)^2-1024 \zeta_3 -640 \zeta_5 +2688 \zeta_7+14336\ .
    \end{split}
\end{equation}

In \cite{Giombi:2017cqn} the AdS$_{2}$/CFT$_{1}$ correspondence was used to find the anomalous dimension of such an insertion at strong coupling via a Witten diagram calculation. On the string theory side it corresponds to the two-point correlator of a singlet operator $y^a y^a$ (with bare dimension $2$), formed of five $S^{5}$ fluctuations orthogonal to the defect. 

We are able to solve the QSC numerically at finite values of the coupling with very high precision, transitioning smoothly between the weak and strong coupling regimes. Our numerical data is presented in figure \ref{fig:num}. It allows us to substantiate the result of \cite{Giombi:2017cqn}. By fitting the numerical data obtained we can predict\footnote{The leading term in equation \eqref{eqn:strong} matches a fit of the numerical data with a relative error $\sim 10^{-14}$, with the next three subleading terms having a relative errors $\sim 10^{-12}$, $10^{-10}$, and $10^{-8}$. The number of digits of precision of
the subleading terms increases as we assume that more terms in this equation are exact. In particular, if we assume that the first three orders of this equation are exact, the relative error in the fourth subleading term goes down to $\sim 10^{-11}$. We therefore believe that our expression is exact for the first four orders with a good degree of confidence.} the first three\footnote{After the first version of this paper appeared on the arXiv, we were informed by the authors of \cite{Liendo:2018ukf}, that they were able to reproduce the second subleading coefficient using analytic bootstrap techniques developed in \cite{Liendo:2016ymz}.} subleading orders at strong coupling. We find
\begin{equation}\label{eqn:strong}
    \Delta = 2 - \frac{5}{\sqrt{\lambda}} + \frac{295}{24}\frac{1}{\lambda} - \frac{305}{16}\frac{1}{\lambda^{3/2}} + \mathcal{O}\left(\frac{1}{\lambda^2}\right)\;.
\end{equation}
 \\
 
This paper is organised as follows. In section \ref{sec:QSCgen} we review the QSC applied to the ground state of the cusp. In section \ref{sec:qscCusp} we review the discussion of the excited states in \cite{Cavaglia:2018lxi}, allowing us to describe the parallel insertions with QSC methods. In section \ref{sec:num} we present our numerical results for the first excited state for various values of $\phi$ and $\theta$ at finite coupling. We show that these results allow us to successfully interpolate between perturbative calculations at weak and strong coupling. In section \ref{sec:analytic} we describe the analytical solution at weak coupling. We end in section \ref{sec:Conclusions} with a brief discussion.

\section{General aspects of QSC}
\label{sec:QSCgen}

The QSC formalism was originally developed to capture the spectrum of local single-trace operators in the planar limit of $\mathcal{N}=4$ SYM \cite{Gromov:2013pga, Gromov:2014caa}. Pedagogical introductions are available in \cite{Gromov:2017blm, Kazakov:2018hrh, Levkovich-Maslyuk:2019awk}. While this formalism is not naively expected to apply to the cusped Wilson line, it has been shown in \cite{Gromov:2015dfa} that only the asymptotics of the Q-functions, as well as the gluing equations, need to be modified --- while keeping the structure of the QQ-relations unchanged --- in order to obtain the spectrum of the cusped Wilson line.

\subsection{Q-functions and QSC equations}
\label{sec:QSCeqs}
In the following we will denote shifts in the spectral parameter $u$ by the standard notation used in this context, i.e.~for a function $f(u)$ dependent on $u$ we define
\begin{equation}
    f^\pm=f(u \pm i/2)\; ,\quad  f^{[n]}=f(u+n i/2)\;.
\end{equation}
The main set of Q-functions is given by
\begin{equation}
\label{eq:Qfunctionsgen}
    \begin{aligned}
        &\tbf{P}^a(u)\;,\; \tbf{P}_a(u)\;, \quad &a&=1,\dots,4\;,\\
        &\tbf{Q}^i(u)\;,\; \tbf{Q}_i(u)\;, \quad &i&=1,\dots,4\;,\\
        &Q_{a|i}(u)\;, \quad &a&=1,\dots,4;\; i=1,\dots,4\;,
    \end{aligned}
\end{equation}
which are sufficient for all practical purposes. In fact, the eight $\Ps$-functions form a basis for the whole Q-system, where all other Q-functions can be obtained using the QQ-relations.

The set of single-index Q-functions \eqref{eq:Qfunctionsgen} is analytic in the complex $u$-plane, apart from prescribed branch points and cuts. The functions $\tbf{P}_a(u)$ have a single branch cut\footnote{As this branch cut does not run through infinity, it is called a short cut.} between $u=\pm 2g$\footnote{$g=\frac{\sqrt{\lambda}}{4 \pi}$, where $\lambda$ is  the 't Hooft coupling.}. Similarly, the functions $\tbf{Q}_i(u)$ have an infinite tower of branch cuts: between $u=\pm 2g$ on the real axis, and its copies, shifted by integer multiples of $i$ into the lower half-plane. The single short branch cut of the $\Ps$-functions can be resolved by the Zhukovsky variable $x(u)$, defined as
\begin{equation}
\label{eq:zhuk}
    x(u)= \frac{u}{2g} + \sqrt{\br{\frac{u}{2g}}^2-1}\;.
\end{equation}

As stated above, the Q-functions satisfy a number of relations, called QQ-relations. First, $Q_{a|i}$ is found through the finite-difference equation
\begin{equation}
\label{eq:FinDiff}
    Q_{a|i}^+ - Q_{a|i}^-=\tbf{P}_a \tbf{Q}_i\;.
\end{equation}
Additionally, the Q-functions have to satisfy
\begin{equation}
\label{eq:PQfromPQQai}
    \begin{split}
         \tbf{Q}_i=-\tbf{P}^a Q_{a|i}^\pm\;,\\
         \tbf{P}_a=-\tbf{Q}^i Q_{a|i}^\pm\;,
    \end{split}
\end{equation}
while the single-index Q-functions obey
\begin{equation}
    \textbf{P}^a \textbf{P}_a = \textbf{Q}^i \textbf{Q}_i =0\;.
\end{equation}
Similar relations to the ones above hold for all indices raised/lowered, with the additional property
\begin{equation}
    \begin{split}
         Q_{a|i}Q^{a|j}=-\delta_i^j\;,\\
         Q_{a|i}Q^{b|i}=-\delta_a^b\;.
    \end{split}
\end{equation}
The above set of equations is known as QQ-relations. In order to find the physical spectrum, the QQ-relations have to be supplementend with the so-called gluing conditions. We describe the relevant gluing conditions below.

\subsection{QSC for cusped Wilson line}
\label{sec:groundCusp}
It was shown in \cite{Gromov:2015dfa} that the QSC can be used to capture the spectrum of a cusped Wilson line with orthogonal scalar insertions at the cusp. This requires two modifications: First, the asymptotics of the Q-functions need to be modified, while the QQ-relations \eqref{eq:FinDiff} and \eqref{eq:PQfromPQQai} are unchanged. Second, the gluing equations need to be changed w.r.t.~to those used for the local operators as we describe below. While for a generic state in $\mathcal{N}=4$ SYM the Q-functions with raised indices can be interpreted as the Hodge-duals of Q-functions with lower indices, they satisfy a simple relation\footnote{This also holds in the left-right symmetric subsector of $\mathcal{N}=4$ SYM, which contains the $sl(2)$-sector.} in the case of the cusped Wilson line:
\begin{equation}
\label{eq:RaiseLowerIndices}
    \begin{split}
        \tbf{P}^a(u)=\chi^{ab}\tbf{P}_b(u)\;,\\
        \tbf{Q}^i(u)=\chi^{ij}\tbf{Q}_j(u)\;,
    \end{split}
\end{equation}
with the constant matrix $\chi$ given by
\begin{equation}
\label{eq:Chi}
\chi^{ab}=\chi^{ij}=
    \begin{pmatrix}
        0 & 0 & 0 & -1\\
        0 & 0 & 1 & 0\\
        0 & -1 & 0 & 0\\
        1 & 0 & 0 & 0\\
    \end{pmatrix}.
\end{equation}

For generic angles $\phi, \theta$, the asymptotics of the single-index Q-functions for the cusp without operator insertions\footnote{This corresponds to $L=0$ in the conventions of \cite{Gromov:2015dfa}.} are given by
\begin{equation}
\label{eq:PsAsym}
    \begin{aligned}
        \Ps_{1}(u) &\simeq C \epsilon^{1/2}\, u^{-1/2}\, e^{+\theta u}f(+u)\;, &\Qs_{1}(u) &\simeq C \epsilon'^{1/2+\Delta}\, u^{-1/2}\, e^{+\phi u}F(+u)\;, \\
        \Ps_{2}(u) &\simeq  C \epsilon^{1/2}\, u^{-1/2}\, e^{-\theta u}f(-u)\;, & \Qs_{2}(u) &\simeq C \epsilon'^{1/2+\Delta}\, u^{-1/2}\, e^{-\phi u}F(-u)\;,\\
        \Ps_{3}(u) &\simeq  \frac{1}{C} \epsilon^{3/2}\, u^{+3/2}\, e^{+\theta u}g(+u)\;,& \Qs_{3}(u) &\simeq \frac{1}{C} \epsilon'^{3/2-\Delta}\, u^{+3/2}\, e^{+\phi u}G(+u)\;,\\
        \Ps_{4}(u) &\simeq  -\frac{1}{C} \epsilon^{3/2}\, u^{+3/2}\, e^{-\theta u}g(-u)\;, & \Qs_{4}(u) &\simeq -\frac{1}{C} \epsilon'^{3/2-\Delta}\, u^{+3/2}\, e^{-\phi u}G(-u)\;,
    \end{aligned}
\end{equation}
where the arbitrary constant $C$ can be set to $1$, and where
\begin{equation}
\begin{aligned}
\label{eq:FsandGs}
    f(u) &= 1 + \sum_{n=1}^{\infty}\frac{a_{n}}{u^{n}}\;,\quad & F(u) &= 1 + \sum_{i=n}^{N}\frac{c_{i}}{u^{i}}\;,\\
    g(u) &= 1 + \sum_{n=1}^{\infty}\frac{b_{n}}{u^{n}}\;,\quad & G(u) &= 1 + \sum_{i=n}^{N}\frac{d_{i}}{u^{i}}\; .
\end{aligned}
\end{equation}
Due to the half-integer powers of $u$ in both the $\Pad$ and $\Qid$, it is often convenient to define
\begin{equation}
\label{eq:PQtopq}
    \ps_a(u)=\frac{\Ps_a(u)}{u^{1/2}}\;, \quad \qs_i(u)=\frac{\Qs_i(u)}{u^{1/2}}\;,
\end{equation}
so that the large-$u$ expansion of the $\pad$ and $\qid$ runs in integer powers of $u$. The $\pad$ can be efficiently represented as an infinite series in inverse powers of  the Zhukovsky variable $x(u)$ \eqref{eq:zhuk}.

In the above expressions the parameter
$\epsilon$ and the combination $a_1-b_1$
are not fully independent already at the level of the
QQ-relations. They can be expressed in terms of the angles $\phi,\, \theta$, and $\Delta$, while $\Delta$ can be expressed in terms of the angles and the first three subleading expansion coefficients in \eqref{eq:FsandGs}. The detailed results can be found in \cite{Gromov:2015dfa}, where they were derived by expanding the Baxter equation \eqref{eq:Baxgen} to fourth order in large $u$. However, we found that the same results can be obtained by using a novel set of relations, called $\PP{}{} \QQ{}{}$-relations, introduced in section \ref{sec:PPQQ}. By expanding the first four of them to fourth order in large $u$, we reproduce the same constraints in a computationally easier way. The specific $\PP{}{}\QQ{}{}$-relations used are listed in eq.~\eqref{eq:PQExpand}.

\subsection{Gluing conditions}
The QQ-relations do not fix all parameters in the expansions \eqref{eq:PsAsym}, so that further relations are needed to obtain a closed system of equations, determining the spectrum of operators. This can generally be achieved by relating some Q-functions to their analytic continuations through the branch cut on the real axis, denoted by $\tilde{\text{Q}}$. Historically, there were two ways to achieve this: the $\Ps \mu$-system and the $\Qs \omega$-system. They relied on constructing antisymmetric matrices $\mu$ --- relating $\Ps$ and $\Pst$ --- or $\omega$ --- relating $\Qs$ and $\Qst$. The remaining parameters were fixed by the fact that $\mu$ and $\omega$ can be expressed as combinations of $\Ps$ and $\Pst$, or $\Qs$ and $\Qst$, respectively.

A simpler construction for local single-trace operators was proposed in \cite{Gromov:2015vua}, where the gluing conditions are given by
\begin{equation}
    \Qst_1=\alpha_1 \conj{\Qs}^2\ , \quad \Qst_2=\alpha_2 \conj{\Qs}^1\ , \quad \Qst_3=\alpha_3 \conj{\Qs}^4\ , \quad \Qst_4=\alpha_4 \conj{\Qs}^3\ ,
\end{equation}
where the $\alpha_i$ are some constants, and the bar denotes complex conjugation. In the case of the cusped Wilson line, these conditions are adapted to \cite{Gromov:2015dfa}
\begin{equation}
\label{eqn:gluground}
    \begin{pmatrix}
        \qst_1(u)\\
        \qst_2(u)\\
        \qst_3(u)\\
        \qst_4(u)
    \end{pmatrix}=\begin{pmatrix}
        1 & 0 & 0 & 0\\
        0 & 1 & 0 & 0\\
        0 & \alpha_1 \sinh{(2 \pi u)} & 1 & 0\\
        \alpha_2 \sinh{(2 \pi u)} & 0 & 0 & 1
    \end{pmatrix}
    \begin{pmatrix}
        \qs_1(-u)\\
        \qs_2(-u)\\
        \qs_3(-u)\\
        \qs_4(-u)
    \end{pmatrix},
\end{equation}
where $\alpha_1$ and $\alpha_2$ are two complex parameters.

\subsection{Baxter equation}
It is often convenient to rewrite the QQ-relations in section \ref{sec:QSCeqs} in a way as to eliminate $Q_{a|i}$. Doing so, one obtains the so-called Baxter equation for the functions $\tbf{Q}_i$, with coefficients dependent on the functions $\tbf{P}^a$ and $\tbf{P}_a$. This was first done in \cite{Alfimov:2014bwa}, with the result being
\begin{equation}
\label{eq:Baxgen}
    D_0 \tbf{Q}^{[+4]} - \left[ D_1 -\tbf{P}_a^{[+2]}\tbf{P}^{a[+4]}D_0 \right]\tbf{Q}^{[+2]}+\frac{1}{2}\left[ D_2 - \tbf{P}_a\tbf{P}^{a[+2]} D_1 +\tbf{P}_a\tbf{P}^{a[+4]} D_0 \right]\tbf{Q} +\text{c.c.}=0.
\end{equation}
The four solutions to this fourth-order finite-difference equation are the four functions $\tbf{Q}_i$. A shorter, simple derivation of this result and an explicit form of the determinants $D_i$ is presented in appendix \ref{sec:AppendixBaxter}.

\subsection{$\mds{P} \mds{Q}$-relations}
\label{sec:PPQQ}
As will be seen later, the Baxter equation can be useful to derive constraints on the Q-functions. However, there is a different set of equations  depending again only on the $\Ps$- and $\Qs$-functions, which are algebraically simpler, but have no free index. These are called the $\mds P \mds Q$-relations, and can be derived from the QQ-relations in section \ref{sec:QSCeqs}. We introduce the notation
\begin{equation}
    \PP{m}{n}=\Ps_a^{[+m]}\Ps^{a [+n]}\;, \quad \QQ{m}{n}=\Qs_i^{[+m]}\Qs^{i [+n]}\;.
\end{equation}
Using the QQ-relations \eqref{eq:FinDiff} and \eqref{eq:PQfromPQQai}, we find
\begin{equation}
\label{eq:PPQQmixed}
    \QQ{2n}{0}=\PP{0}{2n}-\sum_{m=1}^{n-1}\PP{2m}{2n}\QQ{2m}{0}
\end{equation}
for $n \in \mathbb{N}$, which can be rewritten in terms of products of just $\Qs$-functions or $\Ps$-functions. Given a set of numbers $c=\{c_1,\dots,c_{l_c}\}$ of length $l_c$, and defining
\begin{equation}
    \PP{}{}(c)=\prod_{i=1}^{l_c-1}\PP{c_i}{c_{i+1}}\ ,
\end{equation}
we find
\begin{equation}
\label{eq:PPQQpure}
    \QQ{2n}{0}=\sum_{c}(-1)^{l_c}\PP{}{}(c)\;,
\end{equation}
where the sum runs over all ordered sets with unique even entries, such that the first entry is $0$, and the last is $2n$. For a derivation of eqs.~\eqref{eq:PPQQmixed} and \eqref{eq:PPQQpure} see appendix \ref{sec:AppendixPPQQ}.

As described in section \ref{sec:groundCusp}, these relations provide a computationally efficient method to impose constraints on the expansion parameters of the Q-functions. To derive these constraints, the first four $\PP{}{}\QQ{}{}$-relations, given by\footnote{Here we have used eqs.~\eqref{eq:RaiseLowerIndices} and \eqref{eq:Chi}.}
\begin{equation}
\label{eq:PQExpand}
    \begin{split}
        0&=\QQ{0}{2}+\PP{0}{2}\;,\\
        0&=\QQ{0}{4}+\PP{0}{4}-\PP{0}{2}\PP{2}{4}\;,\\
        0&=\QQ{0}{6}+\PP{0}{6}-\PP{0}{2}\PP{2}{6}-\PP{0}{4}\PP{4}{6}+\PP{0}{2}\PP{2}{4}\PP{4}{6}\;,\\
        0&=\QQ{0}{8}+\PP{0}{8}-\PP{0}{2}\PP{2}{8}-\PP{0}{4}\PP{4}{8}-\PP{0}{6}\PP{6}{8}\\
        &+\PP{0}{2}\PP{2}{4}\PP{4}{8}+\PP{0}{2}\PP{2}{6}\PP{6}{8}+\PP{0}{4}\PP{4}{6}\PP{6}{8}-\PP{0}{2}\PP{2}{4}\PP{4}{6}\PP{6}{8}\;,
    \end{split}
\end{equation}
are expanded to fourth order at large $u$.

\section{The excited cusp}\label{excitedcusp}
\label{sec:qscCusp}
In this section we find the anomalous dimension of a single {\it parallel insertion} 
at the cusp of the Wilson line (see eq.~\eqref{eqn:cuspdef}) using the QSC, and describe the details of the QSC setup to do this. First, we review the ladders limit in which the calculation simplifies, which originally appeared in \cite{Cavaglia:2018lxi} and where the excited states of the cusp where first explored. Then, we go beyond the ladders limit and develop a QSC description of the problem in full generality. Finally, we take the limit of a straight Wilson line with a single insertion $\PhiDotn$, which is equivalent to calculating its spectrum in a one-dimensional CFT. We present the asymptotics and gluing conditions in this case. 

\subsection{The ladders limit} \label{sec:lad}

\begin{figure}
    \centering
    \includegraphics[scale=0.85]{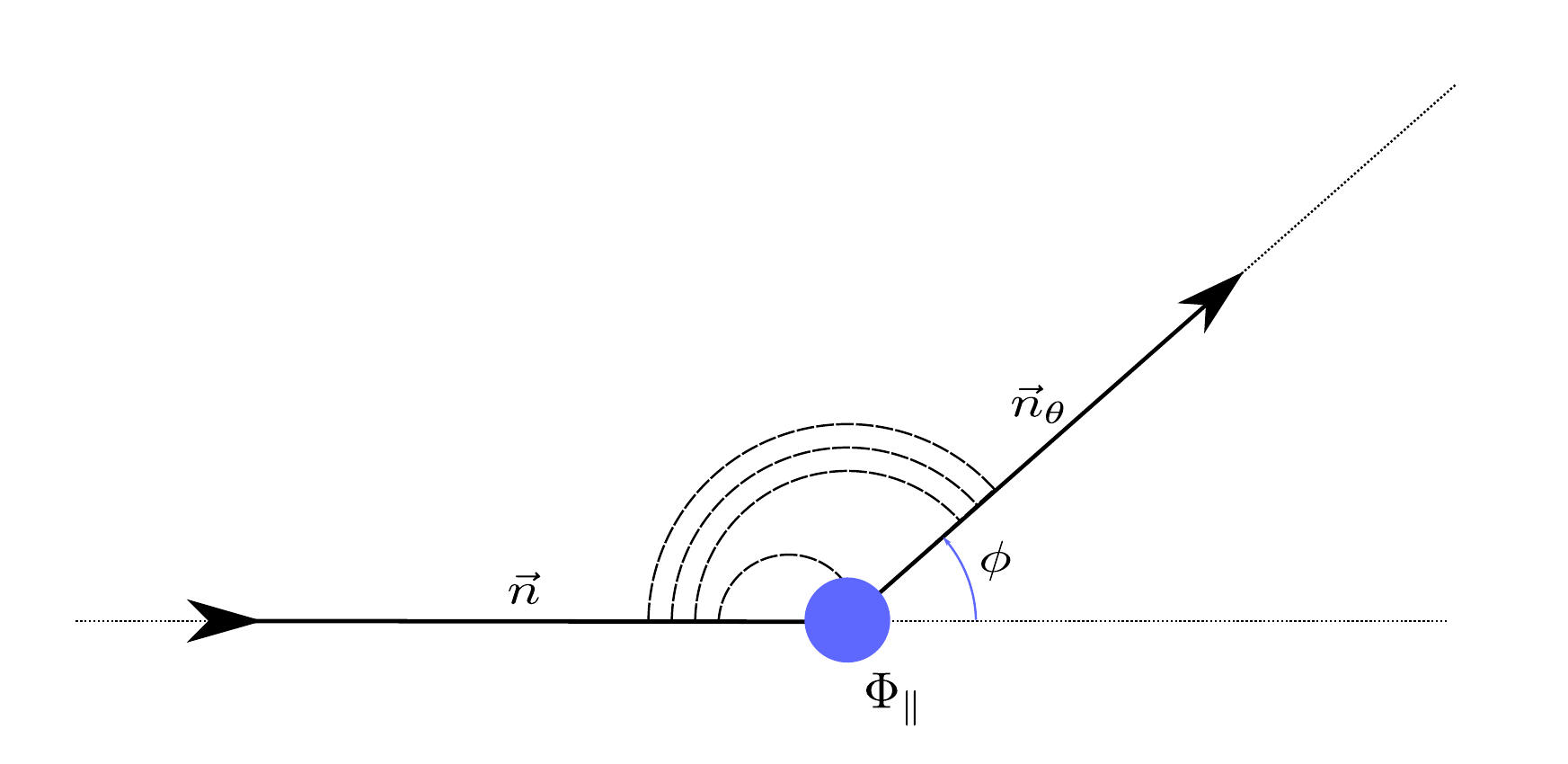}
    \caption{Ladders limit: Dashed lines represent scalar propagators. Each scalar propagator joining two Wilson rays contains a factor of $g^2 \cos\theta$. The blue dot is a scalar insertion of $\Phi_{\parallel}$ at the cusp.}
    \label{fig:ladders}
\end{figure}

The angle between the unit vectors coupling to the scalars on the two Wilson rays is $\theta$. Consequently, propagators connecting scalars on the two rays of the Wilson line in figure \ref{fig:ladders} contain a factor of $g^2 \cos \theta$. The ladders limit \cite{Erickson:1999qv,Erickson:2000af,Correa:2012nk} is obtained by taking the coupling $g\rightarrow 0$ and $\theta\rightarrow i\infty$, in such a way that $\hat{g} = \frac{g}{2} e^{-i \theta/2}$ is kept constant. In this limit, only Feynman diagrams that contain the highest power of $\cos\theta$ survive. The Feynman diagrams at loop order $L$ correspond to ladder diagrams, that is, diagrams that contain $L$ scalar propagators beginning on one of the Wilson lines and ending on the other, see figure \ref{fig:ladders}.

In this limit, the QSC simplifies considerably.
Most of the coefficients in \eq{eq:FsandGs} become suppressed by powers of $g$. As a result the $\Pad$ become explicit simple functions, and therefore the coefficients of the Baxter equation \eq{eq:Baxgen} are quite simple \cite{Cavaglia:2018lxi}. Additionally, the fourth order Baxter equation factorises into two second order ones. In \cite{Cavaglia:2018lxi}, the following second order Baxter equation was obtained:
\begin{align}
\label{eqn:baxlad}
(4\hat g^2+2 \Delta u \sin\phi - 2 u^2 \cos\phi)\tbf q (u)  + u^2 \tbf q(u - i) + u^2 \tbf q(u + i) = 0\;.
\end{align}
The two remaining ${\bf q}$-functions can be found by replacing $\Delta\to-\Delta$. The Baxter equation admits solutions for generic values of $\Delta$. In order to find the physical spectrum, one needs to introduce an appropriate quantisation condition, which constrains the allowed values of $\Delta$ to a discrete set corresponding to physical operators. Thus, it plays the same role as the gluing conditions. In the ladders limit the quantisation condition is given by \cite{Cavaglia:2018lxi}
\begin{equation}
    \Delta=-\frac{2 \hat{g}}{\sin{\phi}}\frac{q_+(0)\Bar{q}'_+(0)+\Bar{q}_+(0)q'_+(0)}{q_+(0)\Bar{q}_+(0)}\, ,
\end{equation}
where $q_+$ denotes the solution of the Baxter equation \eqref{eqn:baxlad} which scales as $q_+ \sim e^{u \phi} u^\Delta$, bar denotes complex conjugation, and the prime denotes the derivative in $u$. As the function $q_+$ contains $\Delta$ in a non-linear way due to the Baxter equation, the quantisation condition allows for many solutions. The physical spectrum at zero coupling $\hat{g}=0$ is given by non-negative integer values $\Delta_0=L$. The ground state solution corresponds to $\Delta_0=0$, while solutions with $\Delta_0=L>0$ are interpreted as excited states, and correspond insertions of
a particular combination of $L$
scalars from $\Phi_\parallel$. As the value of the coupling increases, the dimension $\Delta$ of every excited state splits into two. The first excited state, i.e.~the case with $L = 1$, is considered in this paper and is depicted in figure \ref{fig:cusp}. In the ladders limit, a single insertion of $\Phi_\parallel$ in the point-splitting regularisation scheme can be written explicitly \cite{Cavaglia:2018lxi}.
At weak coupling one of two of the first excited states becomes
\begin{equation}
    \Phi_\parallel\propto \left( \vec{\Phi} \cdot \vec{n}+\vec{\Phi}\cdot\vec{n}_\theta \right)+{\cal O}(g^2)\ ,
\end{equation}
which is the state we will study in this paper. The second state of the cusp with $\Delta_0=1$
should become a descendant of the ground state in the straight line limit and is thus trivial.

\subsection{Beyond the ladders limit}
The ladders limit resums a part of all Feynman diagrams to all orders in perturbation theory in the effective coupling $\hat g$. This results in a non-trivial function $\Delta(\hat g)$. However, the limit involves taking $\theta\to i\infty$, but our goal is to have $\theta$ finite. For finite $\theta$ but small $g$ we still have an explicit form of the $\Pad$, as only finitely many terms survive. We can plug the $\Pad$ into the general form of the Baxter equation \eqref{eq:Baxgen}, resulting in a very complicated expression, which can be provided upon request. In particular, we have not been able to factorise it into two second order equations. Nevertheless, we managed to solve it explicitly as we describe below.

Two solutions turn out to be independent of $\theta$ and can thus be deduced from \eqref{eqn:baxlad}. For the first excited state with $\Delta=1+{\cal O}(g^2)$ we find
\begin{equation}
\tbf q_1(u)=u e^{+u \phi}\;\;,\quad \tbf q_2(u)=u e^{-u \phi}\;.
\end{equation}
The other two solutions have a more complicated singularity structure. In order to describe them in a suitable basis of functions, we recall that at weak coupling the branch cuts of the $\qid$ collapse into poles, so that any solution of the Baxter equation has a prescribed pole structure --- rather than a branch cut structure --- in the lower half plane, spaced out in intervals of $i$. 
When solving the Baxter equation around small coupling it was noticed ~\cite{Leurent:2013mr,Marboe:2014gma,Gromov:2015dfa} that one can restrict
oneself to the basis containing polynomials, shifted inverse powers $\frac{1}{(u+i n)^{a}}$, and certain special functions, called $\eta$-functions, $i$-periodic combinations of $\eta$-functions, and products thereof. We define (generalised) $\eta$-functions as \cite{Gromov:2015dfa}
\begin{align}\label{eqn:etadef}
\eta^{z_{1},z_{2},\cdots,z_{k}}_{s_{1},s_{2},\cdots,s_{k}}(u)\equiv \sum_{n_{1}>n_{2}>\cdots>n_{k}\geq 0}\frac{z_{1}^{n_1}z_{2}^{n_2}\cdots z_{k}^{n_{k}}}{(u + i n_{1})^{s_{1}}(u + i n_{2})^{s_{2}}\cdots(u+i n_{k})^{s_{k}}}\ .
\end{align}
In the case where all the twists $z_{i}=1$, they reduce to the $\eta$-functions that appeared in the weak coupling calculations of \cite{Leurent:2013mr,Marboe:2014gma}. The $\eta$-functions have some algebraic properties that make them particularly amicable for solving finite-difference equations. We have
\begin{align}
\eta_{s_{1},s_{2},\cdots, s_{n},s}^{z_{1},z_{2},\cdots, z_{n},z[+2]} =  \frac{1}{z_{1}z_{2}\cdots z_{n} z}\eta_{s_{1},s_{2},\cdots, s_{n},s}^{z_{1},z_{2},\cdots, z_{n},z} - \frac{1}{z}\frac{1}{u^{s}}\eta_{s_{1},s_{2},\cdots, s_{n}}^{z_{1},z_{2},\cdots, z_{n}}.
\end{align}
This means that we can construct a general ansatz out of $\eta$-functions and rational functions, plug it into the Baxter equation, and then use the above relation to eliminate the respective shifts in the argument. We then demand that the remainder vanishes term by term, thus fixing the parameters of the ansatz. We find
\begin{multline}
e^{-u \phi} \tbf q_3(u)=  \left( u \eta^{\exp(2 i \phi)}_1-\frac{1}{2}(1+i \cot \phi) \right)-\frac{i}{\cos \theta+1}\frac{1}{2}(1+i \cot \phi)\frac{1}{u}+\frac{i}{8}\sec^4\frac{\theta}{2}\cot \frac{\phi}{2}\frac{1}{u^2}\\
-\frac{1}{8}\sec^4\frac{\theta}{2}\cot^2 \frac{\phi}{2} u \eta^1_3-\frac{1}{\cos \theta +1}\cot \frac{\phi}{2}u \eta^{\exp(2 i \phi)}_{2}+\frac{1}{8}\sec^4 \frac{\theta}{2}\cot^2 \frac{\phi}{2} u \eta^{\exp(2 i \phi)}_{3}\ ,
\end{multline}
while the remaining solution $\tbf q_4$ is  obtained by taking $\phi\to-\phi$.

The existence of these solutions for the Baxter equation with general $\theta$ shows that the excited states, found initially in the ladders limit, are still well defined for general $\theta$.
We will also use an explicit form of these $\tbf q$-functions at the leading order in $g$ for the analysis of the untwisting limit when one sends both $\phi,\theta\to 0$. In the next section we consider these limits.

\subsection{QSC for straight line case}

Now we consider the straight line case $\phi\to 0$. We also take
$\theta\to 0$. Note that for the ground state the result is that all ${\bf P}$- and ${\bf Q}$-functions are simply zero, as a consequence of the supersymmetry. This is not the case for the excited state, as the corresponding operators do not preserve any supersymmetries. 

These limits are singular in the natural basis of ``twisted'' Q-functions, i.e. those with ``pure'' large $u$ asymptotic of the type $e^{\alpha u}u^\beta$. In order to obtain the untwisted Q-functions, one should form linear combinations of the twisted Q-functions which have a finite limit. We call this procedure untwisting. 

First, we take $\phi \to 0$, which changes the asymptotics of the $\Qid$. We present them in the case with $\phi = 0$ and nonzero $\theta$ in appendix \ref{apd:untwPh}. Next, we take $\theta\rightarrow 0$. We find the fully untwisted asymptotics of the Q-functions to be as follows:
\begin{equation}
\begin{aligned}
\label{eqn:PandQasym}
\tbf P_{1} &\sim A_{1} u^{3/2} f_{1}(u)\;, & \tbf Q_{1} &\sim B_{1} u^{3/2+\Delta} g_{1}(u)\;, \\
\tbf P_{2} &\sim A_{2} u^{-3/2}f_{2}(u)\;, & \tbf Q_{2} &\sim B_{2} u^{1/2+\Delta} g_{2}(u)\;,\\
\tbf P_{3} &\sim A_{3} u^{1/2} f_{3}(u)\;, & \tbf Q_{3} &\sim B_{3} u^{-3/2-\Delta} g_{3}(u)\;,\\
\tbf P_{4} &\sim A_{4} u^{-5/2} f_{4}(u)\;, & \tbf Q_{4} &\sim B_{4} u^{-5/2-\Delta} g_{4}(u)\;,
\end{aligned}
\end{equation}
where
\begin{equation}
\label{eqn:fandg}
            f_{a}(u) = 1+{\cal O}\left(\frac{1}{u^2}\right)\;, \quad \quad g_{i}(u) = 1+{\cal O}\left(\frac{1}{u^2}\right)\;.
\end{equation}
We notice that the asymptotics of the untwisted functions $\tbf P_{2}$ and $\tbf P_{3}$ differ by an even power of $u$, as do those of $\tbf P_{1}$ and $\tbf P_{4}$. This leaves us with a ``gauge symmetry'', given by the linear transformation
\begin{equation}
\label{eq:PGaugeSymmetry}
\tbf P_{1} \rightarrow \tbf P_{1} + \rho \, \tbf P_{4}\;, \quad \quad \tbf P_{3} \rightarrow \tbf P_{3} + \sigma \, \tbf P_{2}\;,
\end{equation}
which is a freedom of the QSC construction.
For example, one can use this symmetry to fix two of the large $u$ expansion coefficients of the $\Pad$ to be zero, as is done in eq.~\eqref{eqn:cexpn}.
Note that the power structure of the asymptotics of the Q-functions in equation \eqref{eqn:PandQasym} is very similar to those of the Q-functions that describe local single-trace operators in $\mathcal{N} = 4$ SYM. Plugging them into the QQ-relations \eqref{eq:FinDiff} and \eqref{eq:PQfromPQQai}, we can derive constraints on the leading order coefficients. This is done in the standard way, by first finding the asymptotics of $Q_{a|i}$ using eq.~\eqref{eq:FinDiff}, followed by using eq.~\eqref{eq:PQfromPQQai} to obtain a set of independent equations constraining the leading order coefficients. Its solutions are given by
\begin{equation}
\begin{split}
\label{eqn:AABB}
A_{1}A_{4} &= \frac{1}{12} i (\Delta - 1) \Delta (\Delta + 
     3) (\Delta + 4) \; ,\\
A_{2}A_{3} &= -\frac{1}{6}
     i\Delta (\Delta + 1) (\Delta + 
     2) (\Delta + 3)\; ,\\
B_{1}B_{4} &= \frac{i \Delta (\Delta + 1) (\Delta + 
     3) (\Delta + 4)}{2 (\Delta + 2) (2 \Delta + 3)}\; , \\
B_{2}B_{3} &= \frac{i(\Delta - 1) \Delta (\Delta + 
     2) (\Delta + 3)}{2 (\Delta + 1) (2 \Delta + 3)}\; .
     \end{split}
\end{equation}
Again, in analogy to the local operators case, there are no further constraints on the expansion coefficients. Moreover, we can directly use the formula derived in \cite{Gromov:2014caa} (see also \cite{Gromov:2017blm}) for the case of local operators, in order to obtain the constraints in \eqref{eqn:AABB}, as the derivation only relies on the QQ-relations and the absence of twists. Finally, we need the gluing conditions. The gluing matrix is obtained by untwisting the gluing matrix \eqref{eqn:gluground} for the cusp with orthogonal insertions. The fully untwisted gluing matrix is found to be
\begin{align}\label{eqn:glu}\left(
\begin{array}{c} \tilde{\mathbf{q}}_{i}(u) \\
\tilde{\mathbf{q}}_{2}(u) \\
\tilde{\mathbf{q}}_{3}(u) \\
\tilde{\mathbf{q}}_{4}(u)\end{array}\right) = \left(
\begin{array}{cccc}
1 & 0 & 0 & 0\\
0 & 1 & 0 & 0\\
\alpha \sinh{(2\pi u)} & 0  & 1 & 0\\
0 & -\alpha\sinh{(2\pi u)} & 0 & 1
\end{array}
\right)\left(
\begin{array}{c}\mathbf{q}_{1}(-u)\\
\mathbf{q}_{2}(-u)\\
\mathbf{q}_{3}(-u)\\
\mathbf{q}_{4}(-u)
\end{array}\right),
\end{align}
where $\alpha$ is a complex-valued constant. 

Having deduced the correct asymptotics and the structure of the gluing matrix, we can directly implement the numerical method of~\cite{Gromov:2015wca, Gromov:2015dfa} as we explain in the next section.

\section{Numerical solution}
\label{sec:num}

The asymptotic expansions from section \ref{sec:qscCusp} are used to construct a numerical solution of the QSC. The procedure we follow is essentially the same as the one developed in \cite{Gromov:2015wca}, which was applied to the case of the cusp with orthogonal insertions in \cite{Gromov:2015dfa}. We briefly review the procedure below. 

The first step is to reparametrise the $\Pad$ in terms of the Zhukovsky variable $x(u)$ defined in eq.~\eqref{eq:zhuk}.
After extracting the asymptotics, we get
\begin{align}\label{eqn:zhukcoef}
f_{a}(u) = 1 + \sum_{n = 1}^{M}\frac{c_{a,n}}{x^{2n}} \ ,
\end{align}
where $M$ is some suitably large cutoff. We can now approximate the $\Pad$ everywhere in the complex plane, within some small error due to the finite cutoff $M$. The QQ-relations \eqref{eq:FinDiff} and \eqref{eq:PQfromPQQai} are combined to give
\begin{align}\label{eqn:iter}
\Qaid^{+} - \Qaid^{-} = - \Pad\tbf P^{b}Q_{b|i}^{+}\ ,
\end{align}
which we can use to write a large $u$ asymptotic expansion for $\Qaid$:
\begin{align}
\Qaid\sim u^{N_{a|i}}\sum_{i = 1}^{K}\frac{B_{a|i,n}}{u^{n}}\ .
\end{align}
Here, $N_{a|i}$ is the $(a,i)$-th matrix-element of the the product of the asymptotics of the $\Pad$ and the $\Qid$.  We can define a linear problem for the $B_{a|i,n}$ in terms of the $c_{a,n}$, and obtain $\Qaid$ to arbitrarily high precision by choosing a sufficiently high imaginary part for the initial value of the spectral parameter $u$. Then, using eq.~\eqref{eqn:iter} recursively, we bring the value of the argument close to the branch cut. We proceed by finding the $\Qid$ above the branch cut and $\tilde{\tbf Q}_{i}$ below the branch cut using eq.~\eqref{eq:PQfromPQQai} and
\begin{align}
\label{eq:QtildeFromPtilde}
\tilde{\tbf Q}_{i} = -\tilde{\tbf P}^{a}\Qaid^{+}\ .
\end{align}
Notice that the reparamterisation in terms of $x(u)$ lets us transition from $\Pad$ to $\tilde{\tbf P}_{a}$ by the replacement $x\to 1/x$. All that remains is to impose the gluing conditions \eqref{eqn:glu} in the following optimisation problem. 
We can calculate both sides of eq.~\eqref{eqn:glu} at sampling points $u_{k}$ on the cut (usually the Chebyschev points), and minimise the difference between them. The constant $\alpha$ is given by
\begin{align}
\alpha  =  \frac{\tilde{\tbf q}_{3}(u) - \tbf q_{3}(u)}{\tbf q_{1}(u)\sinh(2\pi u)}= -\frac{\tilde{\tbf q}_{4}(u) - \tbf q_{4}(u)}{\tbf q_{2}(u)\sinh(2\pi u)}\ .
\end{align}
Following \cite{Gromov:2015dfa}, we use
\begin{equation}
    \begin{split}
        F = \sum_{k}&\norm{\qst_{1}(u_{k}) - \qs_{1}(-u_{k})}^{2}+\norm{\qst_{2}(u_{k}) - \qs_{2}(-u_{k})}^{2} \\ &+ \text{Var}\be{\frac{\qst_{3}(u_k) - \qs_{3}(-u_k)}{\qs_{1}(-u_k)\sinh(2\pi u_k)}} + \text{Var}\be{-\frac{\qst_{4}(u_k) - \qs_{4}(-u_k)}{\qs_{2}(-u_k)\sinh(2\pi u_k)}},
    \end{split}
\end{equation}
where Var is the usual variance\footnote{$\text{Var}\be{f_k}=\sum_k \norm{\bar{f}-f_k}^2$, where $\bar{f}$ is the average of all sampled values $f_k$.} of a function, which in this case measures the deviation of $\alpha$ from a constant. The function $F$ is minimised using the Levenberg-Marquardt algorithm as described in \cite{Gromov:2015wca}.

\subsection{Results}
Implementing the numerical method of \cite{Gromov:2015wca} in our case,
we are able to interpolate between the weak and strong coupling regimes with very good precision, see figure \ref{fig:num}. We find a smooth transition between the two regimes, so that our result also serves as a non-trivial check of the AdS$_{5}$/CFT$_{4}$ correspondence.

We can extract a three loop guess for the weak coupling result from the numerical data, which we substantiate by an explicit analytical solution of the QSC at weak coupling in section \ref{sec:analytic}. 
\begin{figure}[t]
    \centering
    \includegraphics[scale=1]{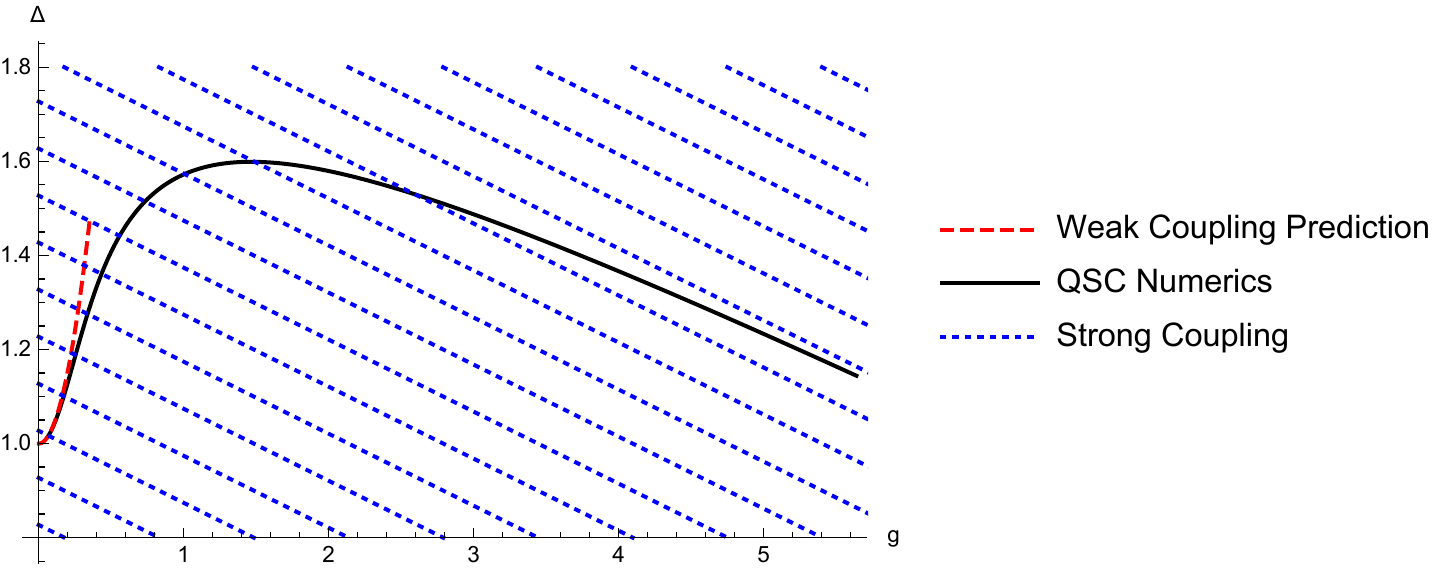}
    \caption{This plot shows our numerical data for the case $\phi = \pi/4$ and $\theta = \pi/8$. The dashed red line represents the one-loop result at weak coupling, which we conjecture to take the form $\Delta(g)=1+4g^2 \cos^2{(\theta/2)}+\mathcal{O}(g^4)$, consistent with both the ladders limit and $\theta \to 0$. The dotted blue lines represent the slope of the classical strong coupling result of \cite{Gromov:2012eu}. The solid black line represents the numerical data obtained with the QSC.}
    \label{fig:phiandthetaalive}
\end{figure}

By making a fit of our numerical data at large $\lambda$ we find that, within the numerical error,
the first four coefficients are given by
\begin{equation}\label{Ds}
    \Delta = 2 - \frac{5}{\sqrt{\lambda}} + \frac{295}{24}\frac{1}{\lambda} - \frac{305}{16}\frac{1}{\lambda^{3/2}} + \mathcal{O}\left(\frac{1}{\lambda^2}\right)\;.
\end{equation}
More precisely, the mismatch with the numerical values we find is 
\begin{align}
    \Delta_{\text{Fit}} - \Delta = -2.7 \times 10^{-14} + \frac{1.5 \times 10^{-11}}{\sqrt{\lambda}} - \frac{3.5 \times 10^{-9}}{\lambda} + \frac{4.1 \times 10^{-7} }{\lambda^{3/2}}+{\cal O}\left(\frac{1}{\lambda^2}\right)\ .
\end{align}
Increasing the precision of our numerical data 
gives a result consistent with \eqref{Ds},
so within reasonable doubt one can assume
\eqref{Ds}
to be exact. 
Assuming this to be correct, we extract 
the following numerical values for the subsequent coefficients:
\begin{align}
    \Delta =2 - \frac{5}{\sqrt{\lambda}} + \frac{295}{24}\frac{1}{\lambda} - \frac{305}{16}\frac{1}{\lambda^{3/2}}  -\frac{19.62538318}{\lambda^{2}} + \frac{259.247338}{\lambda^{5/2}} + \mathcal{O}\left(\frac{1}{\lambda^3} \right)\ .
\end{align}
This is in agreement with the strong coupling result of \cite{Giombi:2017cqn}, given by
\begin{equation}
    \Delta=2-\frac{5}{\sqrt{\lambda}}+\mathcal{O}\left(\frac{1}{\lambda}\right).
\end{equation}

\begin{figure}
    \centering
    \includegraphics[scale=1]{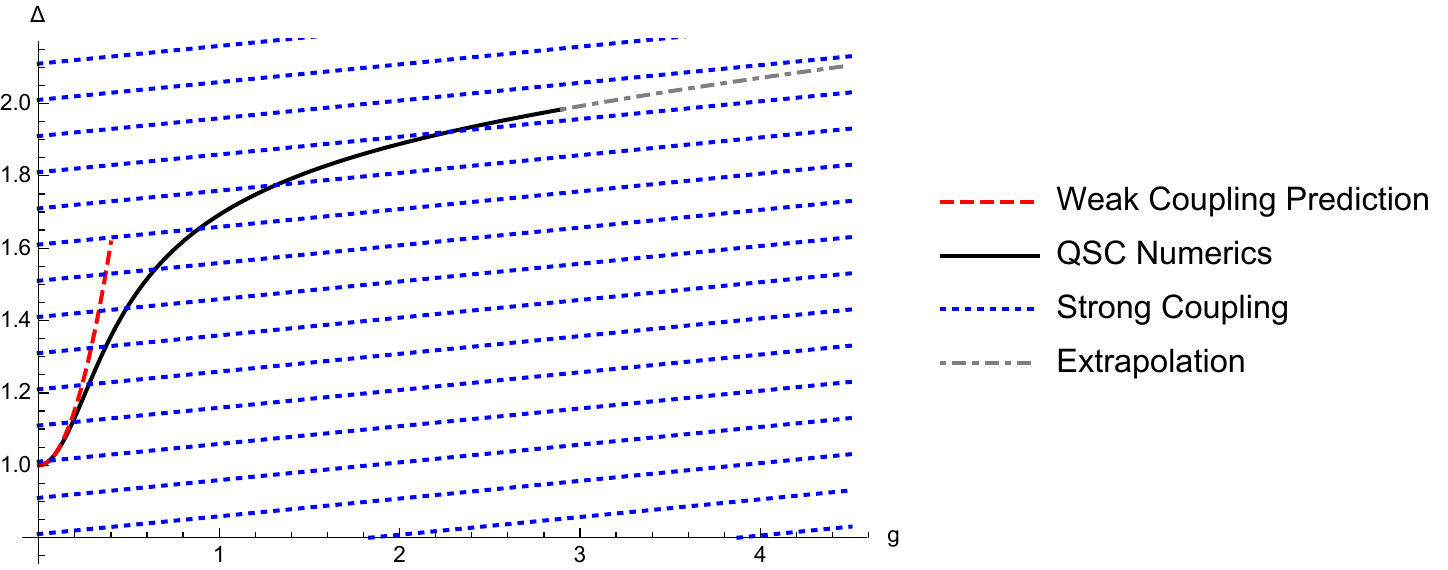}
    \caption{This plot shows our numerical data for the case $\phi = 0$ and $\theta = \pi/8$. The dashed red line represents the one-loop result at weak coupling, which we conjecture to take the form $\Delta(g)=1+4g^2 \cos^2{(\theta/2)}+\mathcal{O}(g^4)$, consistent with both the ladders limit and $\theta \to 0$. The dotted blue lines represent the slope of the classical strong coupling result of \cite{Gromov:2012eu}. The solid black line represents the numerical data obtained using the QSC. The dash-dotted grey line is found by extrapolating the obtained data to higher values of $g$.}
    \label{fig:thetaalive}
\end{figure}

The unknown coefficients could in principle be calculated to arbitrarily high precision, with the only barrier being computational time. Once sufficiently high precision has been reached, one could try to obtain a linear combination of MZVs of various transcendentalities to find analytic predictions.
In appendix \ref{apd:table} we display a table with numerical values for $\Delta$.

We also solve the QSC numerically\footnote{The precision of our numerical data is lower in the cases where at least one twist is present, as compared to when both twists are absent. However, we are still able to reproduce the leading order classical result. At the same time there is no technical difficulty in improving the precision further.} in the case where both $\phi,\, \theta$ are non-zero, and where $\phi$ is zero and $\theta$ is non-zero. Only the classical strong coupling result is known in the case for generic angles \cite{Drukker:2011za, Gromov:2012eu}.

We present our data for $\phi = \pi/4$ and $\theta = \pi/8$ in figure \ref{fig:phiandthetaalive}. Fitting the data at strong coupling, we find
\begin{equation}
    \Delta(\lambda)=-0.0121791 \sqrt{\lambda} +2.0817-\frac{5.45}{\sqrt{\lambda}}+\mathcal{O}\left(\frac{1}{\lambda}\right).
\end{equation}
The numerical prefactor of the linear term in $g$ agrees with the result\footnote{The result in \cite{Gromov:2012eu} is given by a complicated parametric equations. We provide a {\it Mathematica} notebook calculating it to high precision upon request.} of \cite{Gromov:2012eu} to up to seven digits of precision.

For $\phi = 0$ and $\theta = \pi/8$ we obtained numerical results for a wide range in $g$, see figure~\ref{fig:thetaalive}. Fitting the data at strong coupling, we find
\begin{equation}
    \Delta(\lambda)= 0.0038986 \sqrt{\lambda} + 1.9 - \frac{5.}{\sqrt{\lambda}}+\mathcal{O}\left(\frac{1}{\lambda} \right).
\end{equation}
The numerical prefactor of the linear term in $g$ agrees with the result of \cite{Gromov:2012eu} to up to seven digits of precision.

\section{Analytic solution at weak coupling}
\label{sec:analytic}
The QSC can be solved analytically at weak coupling using the iterative procedure developed in \cite{Gromov:2015vua}, which we briefly review below.
\subsection{Revision of method}
\label{sec:analyticalgorithm}

Our starting point is equation \eqref{eqn:iter}. Given the functions $\Ps_a$, and an approximation $Q_{a|i}^{(0)}$ of $Q_{a|i}$ valid up to order $g^{2n}$, the mismatch in eq.~\eqref{eqn:iter} can be expressed as $dS_{a|i}$, i.e.
\begin{equation}
    Q_{a|i}^{(0)+}-Q_{a|i}^{(0)-}+\Ps_a \Ps^b Q_{b|i}^{(0)+}=dS_{a|i}\ ,
\end{equation}
where $dS_{a|i} \sim g^{2n}$ is small. The exact solution can be written as
\begin{equation}
    Q_{a|i}=Q_{a|i}^{(0)}+b_i^{\; j +}Q_{a|j}^{(0)}\ ,
\end{equation}
where the functions $b_i^{\; j}$ can be shown to satisfy the first-order finite-difference equation
\begin{equation}
\label{eq:FinDiffforb}
    b_i^{\;j [+2]}-b_i^{\;j} =-dS_{a|i}Q^{(0)a|j+}+\mathcal{O}\left(g^{4n}\right).
\end{equation}
The advantage of this algorithm is that eq.~\eqref{eq:FinDiffforb} can be solved analytically to high order in $g$, allowing for the procedure to be carried out iteratively. Thus, starting with $Q_{a|i}^{(0)}$ and a weak coupling expansion of the $\Pad$ to sufficiently high order, we are able to find $Q_{a|i}$ to high order in $g$. This allows us to find the weak coupling expansions of $\textbf{q}_i$ and $\tilde{\textbf{q}}_i$ by using eqs.~\eqref{eq:PQfromPQQai} and \eqref{eq:QtildeFromPtilde}. Finally, we impose the gluing conditions \eqref{eqn:glu}. Thus, once we know the Q-functions at tree-level, this algorithm allows us to find the expansion of $\Delta(g)$ for the physical values, as well as the remaining free expansion coefficients of the $\Pad$ in eq.~\eqref{eqn:cexpn}.

\subsection{Q-functions at leading order}
Using the numerical solution of section \ref{sec:num}, we find an ansatz for the scaling of the expansion coefficients of the $\Pad$ in eq.~\eqref{eqn:zhukcoef} at weak coupling. We found the following weak coupling behaviour:
\begin{equation}
\label{eqn:cexpn}
    \begin{split}
        c_{1,1} &= \frac{1}{g^2}\left(c_{1,1,-2} + c_{1,1,0}\ g^2 + c_{2,1,2}\ g^4+\dots\right)\, ,\\
        c_{1,2} &= \frac{1}{g^2}\left(c_{1,2,-2} + c_{1,2,0}\ g^2 + c_{2,2,2}\ g^4+\dots\right)\,,\\
        c_{1,3} &= 0 \text{ (fixed using gauge symmetry)}\,, \\
        c_{1,n>3} &= g^{2n - 6}\left(c_{1,n,2n-6} + c_{1,n,2n-4}\ g^2 + c_{1,n,2n-2}\ g^4+\dots\right)\,,\\
        c_{2,n} &= g^{2n}\left(c_{2,n,2n} + c_{2,n,2n+2}\ g^2 + c_{2,n,2n+4}\ g^4+\dots\right)\,,\\
        c_{3,1} &= \frac{1}{g^2}\left(c_{3,1,-2} + c_{3,1,0}\ g^2 + c_{3,1,2}\ g^4+\dots\right)\,,        \\
        c_{3,2} &= 0 \text{ (fixed using gauge symmetry)}\,, \\
        c_{3,n> 2} &= g^{2n - 4}\left(c_{3,n,2n-4} + c_{3,n,2n-2}\ g^2 + c_{3,n,2n}\ g^4+\dots\right)\,,\\
        c_{4,n} &= g^{2n}\left(c_{4,n,2n} + c_{4,n,2n+2}\ g^2 +         c_{4,n,2n+4}\ g^4+\dots\right)\,.
    \end{split}
\end{equation}
We use this scaling for the $\Pad$ in eqs.~\eqref{eqn:PandQasym} and \eqref{eqn:zhukcoef}, and plug them into the general form of the Baxter equation \eqref{eq:Baxgen}. To leading order in $g$ we obtain
\begin{equation}
\label{eq:BaxterLeadingg}
    \begin{split}
        &+ \br{u+2i}^3 \br{10u^2-10iu+1}q(u+2i)\\
        &- \frac{4\br{10u^6+35iu^5-54u^4-36iu^3-58u^2-96iu+12}}{u+i}q(u+i)\\
        &+2\br{30u^5-57u^3-48u+\frac{32}{u}}q(u)\\
        &- \frac{4\br{10u^6-35iu^5-54u^4+36iu^3-58u^2+96iu+12}}{u-i}q(u-i)\\
        &+ \br{u-2i}^3\br{10u^2+10iu+1}q(u-2i)\\
        &=0.
    \end{split}
\end{equation}
The solutions of this Baxter equation are the $\qid$ at leading order in weak coupling. To find them, we use a general ansatz consisting of the $\eta$-functions defined in eq.~\eqref{eqn:etadef} and powers of $u$. We fix the coefficients of the ansatz by plugging it into eq.~\eqref{eq:BaxterLeadingg}, expanding at large $u$, and solving the resulting expression order by order. In this way we obtain
\begin{equation}
    \begin{split}
        \qs_1&=\frac{4}{3}i u^2 +\mathcal{O}\left( g^2 \right)\ ,\\
        \qs_2&=\frac{12}{5}i g^2 u+\mathcal{O}\left( g^2 \right)\ ,\\
        \qs_3&=-\frac{5}{3}\left( \frac{1}{u^3} -\frac{2i}{u^2}-4i -8u +8 i u^2 \eta^1_2 + 4 i u^2 \eta^1_4 \right)+\mathcal{O}\left( g^2 \right)\ ,\\
        \qs_4&=-3\left( \frac{1}{u^2}-\frac{2i}{u}-2+4i u^2 \eta^1_3 \right)+\mathcal{O}\left( g^2 \right)\ .
    \end{split}
\end{equation}
\subsection{Constructing $Q_{a|i}^{(0)}$}
In order to find the $\qid$ at high order in $g$, we need to use the iteration procedure discussed in section \ref{sec:analyticalgorithm}. Thus, the next step is to find $Q_{a|i}^{(0)}$. Solving equation \eqref{eq:FinDiff} and enforcing equation \eqref{eq:PQfromPQQai}, we find $Q_{a|i}^{(0)-}$, whose non-zero components are given in Appendix \ref{sec:Qai0m}. We note here that there is a remaining gauge symmetry of the $\Qs$-functions, given by
\begin{equation}
    \Qs_1\to \Qs_1+\rho \, \Qs_2\ , \quad \quad \Qs_3 \to \Qs_3 + \sigma\, \Qs_4\ ,
\end{equation}
which can be fixed by imposing that the $\qs$-functions have a large $u$ expansion in $1/u^2$ rather than $1/u$. Having obtained $Q_{a|i}^{(0)-}$, we carry out the procedure in section \ref{sec:analyticalgorithm} to find the functions $\qid$ and $\tilde{\qs}_i$ to high order in $g$. In order to find the physical spectrum we need to impose the gluing conditions \eqref{eqn:glu}, fixing the scaling dimension $\Delta$, as well as the yet unfixed expansion parameters of the $\Pad$ in eq.~\eqref{eqn:cexpn}.

\subsection{Gluing}
After finding the functions $\qid$ and $\tilde{\qs}_i$ to high order in $g$, we impose the gluing conditions \eqref{eqn:glu} in order to obtain the physical spectrum. In order to do this, we consider the scaling behaviour of the $\tilde{\Ps}$-functions used to find the $\tilde{\qs}$-functions. We have
\begin{equation}
    \begin{split}
        \tilde{\Ps}_1&= \frac{1}{g^2}u^3 f_{11}(u)+u f_{12}(u)+ \mathcal{O}\left(g^2\right),\\
        \tilde{\Ps}_2&=\frac{1}{g^4}u^2 f_{21}(u)+\frac{1}{g^2}f_{22}(u)+\frac{1}{u^2}f_{23}(u)+ \mathcal{O}\left(g^2\right),\\
        \tilde{\Ps}_3&=\frac{1}{g^4}u^2+f_{31}(u)+\frac{1}{g^2}f_{32}(u)+\frac{1}{u^2}f_{33}(u)+ \mathcal{O}\left(g^2\right),\\
        \tilde{\Ps}_4&=\frac{1}{g^6}u^3 f_{41}(u)+\frac{1}{g^4}u f_{42}(u)+\frac{1}{g^2}\frac{1}{u}f_{43}(u)+\frac{1}{u^3}f_{44}(u)+ \mathcal{O}\left(g^2\right),
    \end{split}
\end{equation}
where we expanded around small $g$. The functions $f_{ij}(u)$ will depend on the general expansion parameters of the $\Ps$-functions in eq.~\eqref{eqn:cexpn}. For the gluing equations \eqref{eqn:glu} to be consistent, $\alpha$ has to admit a small coupling expansion of the form
\begin{equation}
    \alpha=\alpha(g)=\frac{\alpha_{-6}}{g^6}+\frac{\alpha_{-4}}{g^4}+\dots\ .
\end{equation}
The gluing equations \eqref{eqn:glu} are solved order by order in $g$, starting at $g^{-6}$, by expanding both sides around $u \to 0$. This will result in multiple zeta-functions, arising from the expansion of the $\eta$-functions appearing on both sides. Every order in $g$ fixes a number of coefficients of the $\Ps$-functions, as well as one order in the expansion of $\Delta$ at weak coupling. This procedure was carried out to five loops, with the result summarised in eq.~\eqref{eqn:upto5loops}.
We also verified our analytic result \eq{eqn:upto5loops} numerically with high precision.

\section{Conclusions}
\label{sec:Conclusions}
In this paper we show how the Quantum Spectral Curve~\cite{Gromov:2014caa} description for the cusped Maldacena-Wilson line~\cite{Gromov:2015dfa} can be used to find the first excited state, corresponding to a scalar insertion at the cusp.  More precisely, the scalars inserted are the ones that couple to the Wilson lines. 
Such observables were not previously considered to be accessible with integrability techniques. The integrable description of these states  was first deduced in \cite{Cavaglia:2018lxi} for cusped Wilson lines in the ladders limit. Here we show that this interpretation persists for finite angles and in particular for a straight line.

In the limit where the line becomes straight the equations
for the excited states remain non-trivial, even though the ground state trivialises. We find the spectrum of the first excited state, corresponding to the bare dimension $\Delta_0=1$,
numerically for a wide range of the coupling at high precision. We reproduce previously derived weak and strong coupling results, and interpolate between them. Using the numerical data obtained, we find the first three subleading coefficients in the $1/\sqrt{\lambda}$ expansion at strong coupling. Additionally, we derive an analytic weak coupling expression at five loops using the algorithm of \cite{Gromov:2015vua}, matching our numerical data.

The method presented in this paper can be extended to capture the spectrum of higher excitations. In principle it should be possible to treat the case of very high excitations --- corresponding to a large number of operator insertions --- with a modified version of the Asymptotic Bethe Ansatz.
This direction should be explored further.

One motivation for this study comes from the one-dimensional defect CFT. While no general, non-perturbative integrability method is available to calculate the structure constants in ${\cal N}=4$ SYM, recent investigations show that the Q-functions of the QSC might encode this data \cite{Cavaglia:2018lxi, McGovern:2019sdd, Giombi:2018qox, Giombi:2018hsx, Derkachov:2019tzo, colortwist2}.
The present setup could be an ideal background for this development.

Finally, we would like to highlight that recently, a spin chain description was obtained at weak coupling for scalar insertions in the non-supersymmetric Wilson loop in $\mathcal{N}=4$ SYM \cite{Correa:2018fgz}. The strong coupling result for the expectation of a Wilson loop with scalar insertions was obtained in \cite{Beccaria:2017rbe}. All six scalars of the theory are on equal footing in this case. The existence of a spin chain description at one loop hints that the problem is integrable to all loops, and one could try to apply the QSC method in this case as well.

\section*{Acknowledgments}
We would like to thank A.~Cavagli\`{a}, N.~Drukker, F.~Levkovich-Maslyuk, and A.~Sever for valuable discussions, and in particular F.~Levkovich-Maslyuk for sharing some of his Mathematica notebooks with us. D.G.~was supported by the EPSRC Research Studentship (EP/N509498/1). N.G.~was supported by the STFC grant (ST/P000258/1) and the ERC grant 865075 EXACTC. J.J.~would like to thank H.~Chembati and S.~K.~Nandy for lending us the computing facilities of the CAD Lab at the Indian Institute of Science, where some of the numerical results were obtained.

\appendix

\section{Derivation of Baxter equation}
\label{sec:AppendixBaxter}
In this appendix we want to present a short, modern derivation of the Baxter equation \eqref{eq:Baxgen}, and give an explicit form of the determinants $D_i$. There are multiple ways of deriving the Baxter equation, the first one having been published in \cite{Alfimov:2014bwa}. The easiest derivation the authors are aware of is to start with the trivial $5\times 5$ determinant
\begin{equation}
    \begin{vmatrix}
        \Ps^{a [-4]} & \Ps^{a [-2]} & \Ps^{a} & \Ps^{a [+2]} & \Ps^{a [+4]} & \\
        \Ps^{1 [-4]} & \Ps^{1 [-2]} & \Ps^{1} & \Ps^{1 [+2]} & \Ps^{1 [+4]} & \\
        \Ps^{2 [-4]} & \Ps^{2 [-2]} & \Ps^{2} & \Ps^{2 [+2]} & \Ps^{2 [+4]} & \\
        \Ps^{3 [-4]} & \Ps^{3 [-2]} & \Ps^{3} & \Ps^{3 [+2]} & \Ps^{3 [+4]} & \\
        \Ps^{4 [-4]} & \Ps^{4 [-2]} & \Ps^{4} & \Ps^{4 [+2]} & \Ps^{4 [+4]} & \\
    \end{vmatrix}=0\ ,
\end{equation}
where $a \in \{1,2,3,4\}$. Expanding in the first column, we find
\begin{equation}
\label{eq:BaxPs}
    D_0 \Ps^{a[+4]}-D_1 \Ps^{a[+2]}+D_2 \Ps^{a}-\cD_1 \Ps^{a[-2]}+\cD_0 \Ps^{a[-4]}=0\ ,
\end{equation}
where the determinants $D_i$ are given by
\begin{equation}
    \begin{gathered}
        D_0=
        \begin{vmatrix}
        \Ps^{1[-4]} & \Ps^{1[-2]} & \Ps^{1} & \Ps^{1[+2]}\\
        \Ps^{2[-4]} & \Ps^{2[-2]} & \Ps^{2} & \Ps^{2[+2]}\\
        \Ps^{3[-4]} & \Ps^{3[-2]} & \Ps^{3} & \Ps^{3[+2]}\\
        \Ps^{4[-4]} & \Ps^{4[-2]} & \Ps^{4} & \Ps^{4[+2]}\\
        \end{vmatrix}\ ,
        \quad
        D_1=
        \begin{vmatrix}
        \Ps^{1[-4]} & \Ps^{1[-2]} & \Ps^{1} & \Ps^{1[+4]}\\
        \Ps^{2[-4]} & \Ps^{2[-2]} & \Ps^{2} & \Ps^{2[+4]}\\
        \Ps^{3[-4]} & \Ps^{3[-2]} & \Ps^{3} & \Ps^{3[+4]}\\
        \Ps^{4[-4]} & \Ps^{4[-2]} & \Ps^{4} & \Ps^{4[+4]}\\
        \end{vmatrix}\ ,
        \\
        D_2=
        \begin{vmatrix}
        \Ps^{1[-4]} & \Ps^{1[-2]} & \Ps^{1[+2]} & \Ps^{1[+4]}\\
        \Ps^{2[-4]} & \Ps^{2[-2]} & \Ps^{2[+2]} & \Ps^{2[+4]}\\
        \Ps^{3[-4]} & \Ps^{3[-2]} & \Ps^{3[+2]} & \Ps^{3[+4]}\\
        \Ps^{4[-4]} & \Ps^{4[-2]} & \Ps^{4[+2]} & \Ps^{4[+4]}\\
        \end{vmatrix}\ ,
    \end{gathered}
\end{equation}
while the $\cD_i$ are obtained by inverting the shifts in the corresponding $D_i$. To proceed, we multiply \eqref{eq:BaxPs} with $Q_{a|i}^{-}$, obtaining
\begin{equation}
    D_0 \Ps^{a[+4]}Q_{a|i}^{-}-D_1 \Ps^{a[+2]}Q_{a|i}^{-}+D_2 \Ps^{a}Q_{a|i}^{-}-\cD_1 \Ps^{a[-2]}Q_{a|i}^{-}+\cD_0 \Ps^{a[-4]}Q_{a|i}^{-}=0\ .
\end{equation}
Finally, we use \eqref{eq:FinDiff} and \eqref{eq:PQfromPQQai} to shift $Q_{a|i}$ and contract it with the respective shifted $\Ps_a$. As an example, we obtain for the first term
\begin{equation}
    \begin{split}
         \Ps^{a[+4]}Q_{a|i}^- &= \Ps^{a[+4]}\left[ Q_{a|i}^{+} - \Ps_{a}\Qs_{i} \right]=\Ps^{a[+4]}\left[ Q_{a|i}^{+3}-\Ps_a^{[+2]}\Qs_i^{[+2]}-\Ps_{a}\Qs_{i} \right]\\
         &=-\Qs_i^{[+4]}-\Ps^{a[+4]}\Ps_a^{[+2]}\Qs_i^{[+2]}-\Ps^{a[+4]}\Ps_a \Qs_i\ .
    \end{split}
\end{equation}
Repeating this process for the remaining terms and collecting the different shifts in $\Qs_i$, we obtain \eqref{eq:Baxgen}.

\section{Proof of $\PP{}{} \QQ{}{}$-relations}
\label{sec:AppendixPPQQ}
To derive \eqref{eq:PPQQmixed}, we rewrite $\Qs_i^{[+2n]}$ using \eqref{eq:PQfromPQQai}, shift the resulting $Q_{a|i}^{[+2n-1]}$ to $Q_{a|i}^{+}$ using \eqref{eq:FinDiff}, and contract the resulting expression by again using \eqref{eq:PQfromPQQai}:
\begin{equation}
    \begin{split}
        \QQ{2n}{0}&=\Qs^{i}\Qs_{i}^{[+2n]}=-\Qs^i \Ps^{a[+2n]} Q_{a|i}^{[+2n-1]}=-\Qs^i \Ps^{a[+2n]} \left[Q_{a|i}^+ + \sum_{m=1}^{n-1}\Ps_a^{[+2m]}\Qs_i^{[+2m]} \right]\\
        &=\Ps^{a[+2n]}\Ps_a -\sum_{m=1}^{n-1}\Ps^{a[+2n]}\Ps_a^{[+2m]}\Qs^{i}\Qs_i^{[+2m]}=\PP{0}{2n}-\sum_{m=1}^{n-1}\PP{2m}{2n}\QQ{2m}{0}\ .
    \end{split}
\end{equation}
To derive \eqref{eq:PPQQpure} we will perform a proof by induction. First notice that
\begin{equation}
    \QQ{2}{0}=\PP{0}{2}\ ,
\end{equation}
which is consistent with \eqref{eq:PPQQpure}, as the only set in the sum is $c=\{0,2\}$, thus establishing the inductive hypothesis. From \eqref{eq:PPQQmixed} we find
\begin{equation}
    \begin{split}
        \QQ{2n+2}{0}&=\PP{0}{2n+2}-\sum_{m=1}^{n}\PP{2m}{2n+2}\QQ{2m}{0}=\PP{0}{2n+2}-\sum_{m=1}^n \PP{2m}{2n+2}\sum_{c'}(-1)^{l_{c'}}\PP{}{}\left(c'\right)\\
        &=\sum_{c}(-1)^{l_c}\PP{}{}(c)\ .
    \end{split}
\end{equation}
This derivation merits some explanation: First we use \eqref{eq:PPQQmixed}, followed by using the inductive assumption on $\QQ{2m}{0}$ for every $m\leq n$, so that the sum in $c'$ runs over all ordered sets with unique even entries, with first entry $0$ and last entry $2m$. Finally, we recognise that the resulting expression can be rewritten as the desired result.

\section{Form of Q-functions for $\phi=0$, $\theta \neq 0$}
\label{apd:untwPh}
The $\Pad$ remain unchanged. The asymptotics of the $\Qid$ are given by
\begin{equation}
\begin{aligned}
\tbf Q_{1} &\sim C_{0} \epsilon'^{1/2}_{0} u^{3/2+\Delta} F(+u)\ ,\\
\tbf Q_{2} &\sim C_{0} \epsilon'^{1/2}_{0} u^{1/2+\Delta}F(-u)\ ,\\
\tbf Q_{3} &\sim C_{0} \epsilon'^{3/2}_{0} u^{5/2-\Delta} G(+u)\ ,\\
\tbf Q_{4} &\sim \frac{\Delta-1}{\Delta}C_{0}\epsilon'^{3/2}_{0} u^{3/2-\Delta} G(-u)\ .
\end{aligned}
\end{equation}
Similar to the fully twisted case, the parameters in this expansion satisfy a number of constraints, which can be derived using the $\mathds{P}\mathds{Q}$-relations. We find
\begin{equation}
\epsilon'_{0} = \left(\frac{8{\ }i \sin^4\frac{\theta}{2}}{C_0^{2}(1 - 3 \Delta + 2 \Delta^2)}\right)^{1/2}\ ,
\end{equation}   
and additionally
\begin{equation}
\begin{split}
    a_{1} - b_{1} &= -\frac{\cos(\theta) + 2}{\sin(\theta)}\ ,\\
    a_{1} b_{2} &=  a_{2} \cot(\theta) + a_{1} \csc(\theta)^{2} + 
   a_{1}^{2} \csc(\theta) + a_{2} a_{1} - a_{3} + b_{3}\ .
\end{split}
\end{equation}
Finally, the scaling dimension $\Delta$ satisfies the constraint
\begin{equation}
    \begin{split}
        \br{\Delta-\frac{1}{2}}^2 = \frac{1}{4\br{\cos{\theta}+1}}\Big[ &\br{4 a_2-2b_1^2+2}\cos{2\theta} -4a_2 -24b_1 \sin{\theta}\\ &-4b_1 \sin{2\theta}+2b_1^2 +25 \cos{\theta}+35\Big]\ .
    \end{split}
\end{equation}

\section{Non-zero components of $Q_{a|i}^{(0)-}$}
\label{sec:Qai0m}
\begin{equation}
    \begin{split}
        Q_{1|3}^{(0)-}&=-\frac{5}{3}\ ,\\
        Q_{2|1}^{(0)-}&=\frac{2 u^2}{3}-\frac{2 i u}{3}-\frac{1}{3}\ ,\\
        Q_{2|3}^{(0)-}&=\left(-\frac{20 u^2}{3}+\frac{20 i u}{3}+\frac{10}{3}\right) \eta
   _2^1+\left(-\frac{10 u^2}{3}+\frac{10 i
   u}{3}+\frac{5}{3}\right) \eta _4^1-\frac{20 i
   u}{3}-\frac{10}{3}\ ,\\
   Q_{2|4}^{(0)-}&=\left(-6 u^2+6 i u+3\right) \eta _3^1-3 i\ ,\\
   Q_{3|1}^{(0)-}&=-\frac{4}{3} i u^2 (2 \,c_{3,1,-2}-1)-\frac{8}{3} u
   \,c_{3,1,-2}+\frac{4}{3} i \,c_{3,1,-2}-\frac{4 i
   u^4}{3}-\frac{8 u^3}{3}\ ,\\
   Q_{3|3}^{(0)-}&=\eta _2^1 \left(\frac{40}{3} i u^2 (2
   \,c_{3,1,-2}-1)+\frac{80}{3} u
   \,c_{3,1,-2}-\frac{40}{3} i \,c_{3,1,-2}+\frac{40 i
   u^4}{3}+\frac{80 u^3}{3}\right)\\
   &+\eta _4^1 \left(\frac{20}{3}
   i u^2 (2 \,c_{3,1,-2}-1)+\frac{40}{3} u
   \,c_{3,1,-2}-\frac{20}{3} i \,c_{3,1,-2}+\frac{20 i
   u^4}{3}+\frac{40 u^3}{3}\right)\\
   &-\frac{80}{3} u
   \,c_{3,1,-2}+\frac{10}{3} i (4
   \,c_{3,1,-2}+1)-\frac{40 u^3}{3}+20 i u^2\ ,\\
   Q_{3|4}^{(0)-}&=\eta _3^1 \left(12 i u^2 (2 \,c_{3,1,-2}-1)+24 u
   \,c_{3,1,-2}-12 i \,c_{3,1,-2}+12 i u^4+24
   u^3\right)\\
   &=-3 (4 c_{3,1,-2}+1)-6 u^2+6 i u\ ,\\
   Q_{4|1}^{(0)-}&=\frac{4 u}{3}-\frac{2 i}{3},\\
   Q_{4|2}^{(0)-}&=\frac{3}{5}\ ,\\
   Q_{4|3}^{(0)-}&=\frac{20}{3}\eta _3^1+\frac{5 }{3}\eta _5^1+\left(-\frac{40
   u}{3}+\frac{20 i}{3}\right) \eta _2^1+\left(-\frac{20
   u}{3}+\frac{10 i}{3}\right) \eta _4^1-\frac{40 i}{3}\ ,\\
    Q_{4|4}^{(0)-}&=6\, \eta _2^1+3 \,\eta _4{}^1+(-12 u+6 i)\, \eta _3^1\ ,
    \end{split}
\end{equation}
where $c_{3,1,-2}$ is an expansion parameter in the $\Ps$-functions, see eq.~\eqref{eqn:cexpn}.

\newpage
\section{Numerical data for the first excited state of the straight Wilson line}
\label{apd:table}
We present a table with numerical values of $\Delta(g)$ for the first excited state of the straight Wilson line, for a wide range of values of the coupling. While the data obtained has very high precision, we only display the first ten digits. Data with higher precision is provided upon request.

\begin{table}[h]
\centering
\begin{tabular}{| c | c || c | c || c | c || c | c |} 
\hline
g & $\Delta(g)$ & g & $\Delta(g)$ & g & $\Delta(g)$ & g & $\Delta(g)$ \\  \hline
0.02	&	1.001597441	&	1.06	&	1.685728714	&	2.14	&	1.830067113	&	3.26	&	1.884990678	\\ \hline
0.06	&	1.014193085	&	1.10	&	1.695283395	&	2.18	&	1.832914829	&	3.30	&	1.886303778	\\ \hline
0.10	&	1.038413668	&	1.14	&	1.704278088	&	2.22	&	1.835668877	&	3.34	&	1.887587263	\\ \hline
0.14	&	1.072392557	&	1.18	&	1.712760290	&	2.26	&	1.838333793	&	3.38	&	1.888842123	\\ \hline
0.18	&	1.113587510	&	1.22	&	1.720772316	&	2.30	&	1.840913828	&	3.42	&	1.890069304	\\ \hline
0.22	&	1.159156169	&	1.26	&	1.728351980	&	2.34	&	1.843412965	&	3.46	&	1.891269711	\\ \hline
0.26	&	1.206387712	&	1.30	&	1.735533170	&	2.38	&	1.845834945	&	3.50	&	1.892444210	\\ \hline
0.30	&	1.253054911	&	1.34	&	1.742346346	&	2.42	&	1.848183278	&	3.54	&	1.893593630	\\ \hline
0.34	&	1.297594595	&	1.38	&	1.748818955	&	2.46	&	1.850461270	&	3.58	&	1.894718765	\\ \hline
0.38	&	1.339104679	&	1.42	&	1.754975794	&	2.50	&	1.852672027	&	3.62	&	1.895820376	\\ \hline
0.42	&	1.377215244	&	1.46	&	1.760839321	&	2.54	&	1.854818479	&	3.66	&	1.896899192	\\ \hline
0.46	&	1.411915459	&	1.50	&	1.766429924	&	2.58	&	1.856903387	&	3.70	&	1.897955914	\\ \hline
0.50	&	1.443398234	&	1.54	&	1.771766151	&	2.62	&	1.858929356	&	3.74	&	1.898991211	\\ \hline
0.54	&	1.471948837	&	1.58	&	1.776864913	&	2.66	&	1.860898847	&	3.78	&	1.900005730	\\ \hline
0.58	&	1.497877093	&	1.62	&	1.781741660	&	2.70	&	1.862814186	&	3.82	&	1.901000088	\\ \hline
0.62	&	1.521481720	&	1.66	&	1.786410534	&	2.74	&	1.864677572	&	3.86	&	1.901974881	\\ \hline
0.66	&	1.543034635	&	1.70	&	1.790884507	&	2.78	&	1.866491088	&	3.90	&	1.902930680	\\ \hline
0.70	&	1.562776187	&	1.74	&	1.795175494	&	2.82	&	1.868256708	&	3.94	&	1.903868033	\\ \hline
0.74	&	1.580915745	&	1.78	&	1.799294463	&	2.86	&	1.869976300	&	3.98	&	1.904787470	\\ \hline
0.78	&	1.597634521	&	1.82	&	1.803251526	&	2.90	&	1.871651640	&	4.02	&	1.905689499	\\ \hline
0.82	&	1.613089086	&	1.86	&	1.807056018	&	2.94	&	1.873284412	&	4.06	&	1.906574610	\\ \hline
0.86	&	1.627414836	&	1.90	&	1.810716572	&	2.98	&	1.874876216	&	4.10	&	1.907443273	\\ \hline
0.90	&	1.640729129	&	1.94	&	1.814241184	&	3.02	&	1.876428573	&	4.14	&	1.908295942	\\ \hline
0.94	&	1.653134005	&	1.98	&	1.817637271	&	3.06	&	1.877942930	&	4.18	&	1.909133056	\\ \hline
0.98	&	1.664718494	&	2.02	&	1.820911718	&	3.10	&	1.879420663	&	4.22	&	1.909955035	\\ \hline
1.02	&	1.675560558	&	2.06	&	1.824070933	&	3.14	&	1.880863083	&	4.26	&	1.910762286	\\ \hline
1.06	&	1.685728714	&	2.10	&	1.827120877	&	3.18	&	1.882271441	&	4.30	&	1.911555201	\\ \hline

\end{tabular}
\caption{This table contains the numerical values obtained for $\Delta(g)$ for the first excited state in the limit where both angles $\phi$ and $\theta$ vanish. This corresponds to inserting a single scalar at a point along a straight Wilson line, where the scalar is the same as the ones coupled to the Wilson line. The data obtained is precise to at least 20 digits, the first ten of which are included in the table.}
\end{table}

\bibliographystyle{JHEP.bst}
\bibliography{references}

\providecommand{\href}[2]{#2}\begingroup\raggedright\begin{thebibliography}{10}

\bibitem{Maldacena:1997re}
J.~M. Maldacena, \emph{{The Large N limit of superconformal field theories and
  supergravity}}, \href{https://doi.org/10.1023/A:1026654312961,
  10.4310/ATMP.1998.v2.n2.a1}{\emph{Int. J. Theor. Phys.} {\bfseries 38} (1999)
  1113} [\href{https://arxiv.org/abs/hep-th/9711200}{{\ttfamily
  hep-th/9711200}}].

\bibitem{Witten:1998qj}
E.~Witten, \emph{{Anti-de Sitter space and holography}},
  \href{https://doi.org/10.4310/ATMP.1998.v2.n2.a2}{\emph{Adv. Theor. Math.
  Phys.} {\bfseries 2} (1998) 253}
  [\href{https://arxiv.org/abs/hep-th/9802150}{{\ttfamily hep-th/9802150}}].

\bibitem{Gubser:1998bc}
S.~S. Gubser, I.~R. Klebanov and A.~M. Polyakov, \emph{{Gauge theory
  correlators from noncritical string theory}},
  \href{https://doi.org/10.1016/S0370-2693(98)00377-3}{\emph{Phys. Lett.}
  {\bfseries B428} (1998) 105}
  [\href{https://arxiv.org/abs/hep-th/9802109}{{\ttfamily hep-th/9802109}}].

\bibitem{Beisert:2010jr}
N.~Beisert et~al., \emph{{Review of AdS/CFT Integrability: An Overview}},
  \href{https://doi.org/10.1007/s11005-011-0529-2}{\emph{Lett. Math. Phys.}
  {\bfseries 99} (2012) 3} [\href{https://arxiv.org/abs/1012.3982}{{\ttfamily
  1012.3982}}].

\bibitem{Dorey:2019gkd}
P.~Dorey, G.~Korchemsky, N.~Nekrasov, V.~Schomerus, D.~Serban and
  L.~Cugliandolo, eds., \emph{{Integrability: From Statistical Systems to Gauge
  Theory}}, vol.~106 of \emph{Lecture Notes of the Les Houches Summer School}.
  Oxford University Press, 2019.

\bibitem{Lipatov:1993yb}
L.~N. Lipatov, \emph{{Asymptotic behavior of multicolor QCD at high energies in
  connection with exactly solvable spin models}}, {\emph{JETP Lett.} {\bfseries
  59} (1994) 596} [\href{https://arxiv.org/abs/hep-th/9311037}{{\ttfamily
  hep-th/9311037}}].

\bibitem{Faddeev:1994zg}
L.~D. Faddeev and G.~P. Korchemsky, \emph{{High-energy QCD as a completely
  integrable model}},
  \href{https://doi.org/10.1016/0370-2693(94)01363-H}{\emph{Phys. Lett.}
  {\bfseries B342} (1995) 311}
  [\href{https://arxiv.org/abs/hep-th/9404173}{{\ttfamily hep-th/9404173}}].

\bibitem{Minahan:2002ve}
J.~A. Minahan and K.~Zarembo, \emph{{The Bethe ansatz for N=4
  superYang-Mills}},
  \href{https://doi.org/10.1088/1126-6708/2003/03/013}{\emph{JHEP} {\bfseries
  03} (2003) 013} [\href{https://arxiv.org/abs/hep-th/0212208}{{\ttfamily
  hep-th/0212208}}].

\bibitem{Gromov:2013pga}
N.~Gromov, V.~Kazakov, S.~Leurent and D.~Volin, \emph{{Quantum Spectral Curve
  for Planar $\mathcal{N} = 4$ Super-Yang-Mills Theory}},
  \href{https://doi.org/10.1103/PhysRevLett.112.011602}{\emph{Phys. Rev. Lett.}
  {\bfseries 112} (2014) 011602}
  [\href{https://arxiv.org/abs/1305.1939}{{\ttfamily 1305.1939}}].

\bibitem{Gromov:2014caa}
N.~Gromov, V.~Kazakov, S.~Leurent and D.~Volin, \emph{{Quantum spectral curve
  for arbitrary state/operator in AdS$_{5}$/CFT$_{4}$}},
  \href{https://doi.org/10.1007/JHEP09(2015)187}{\emph{JHEP} {\bfseries 09}
  (2015) 187} [\href{https://arxiv.org/abs/1405.4857}{{\ttfamily 1405.4857}}].

\bibitem{Gromov:2017blm}
N.~Gromov, \emph{{Introduction to the Spectrum of $N=4$ SYM and the Quantum
  Spectral Curve}},  \href{https://arxiv.org/abs/1708.03648}{{\ttfamily
  1708.03648}}.

\bibitem{Kazakov:2018hrh}
V.~Kazakov, \emph{{Quantum Spectral Curve of $\gamma$-twisted ${\cal N}=4$ SYM
  theory and fishnet CFT}},  \href{https://arxiv.org/abs/1802.02160}{{\ttfamily
  1802.02160}}.

\bibitem{Levkovich-Maslyuk:2019awk}
F.~Levkovich-Maslyuk, \emph{{A review of the AdS/CFT Quantum Spectral Curve}},
  \href{https://arxiv.org/abs/1911.13065}{{\ttfamily 1911.13065}}.

\bibitem{Gromov:2015wca}
N.~Gromov, F.~Levkovich-Maslyuk and G.~Sizov, \emph{{Quantum Spectral Curve and
  the Numerical Solution of the Spectral Problem in AdS5/CFT4}},
  \href{https://doi.org/10.1007/JHEP06(2016)036}{\emph{JHEP} {\bfseries 06}
  (2016) 036} [\href{https://arxiv.org/abs/1504.06640}{{\ttfamily
  1504.06640}}].

\bibitem{Maldacena:1998im}
J.~M. Maldacena, \emph{{Wilson loops in large N field theories}},
  \href{https://doi.org/10.1103/PhysRevLett.80.4859}{\emph{Phys. Rev. Lett.}
  {\bfseries 80} (1998) 4859}
  [\href{https://arxiv.org/abs/hep-th/9803002}{{\ttfamily hep-th/9803002}}].

\bibitem{Erickson:2000af}
J.~K. Erickson, G.~W. Semenoff and K.~Zarembo, \emph{{Wilson loops in N=4
  supersymmetric Yang-Mills theory}},
  \href{https://doi.org/10.1016/S0550-3213(00)00300-X}{\emph{Nucl. Phys.}
  {\bfseries B582} (2000) 155}
  [\href{https://arxiv.org/abs/hep-th/0003055}{{\ttfamily hep-th/0003055}}].

\bibitem{Drukker:2000rr}
N.~Drukker and D.~J. Gross, \emph{{An Exact prediction of N=4 SUSYM theory for
  string theory}}, \href{https://doi.org/10.1063/1.1372177}{\emph{J. Math.
  Phys.} {\bfseries 42} (2001) 2896}
  [\href{https://arxiv.org/abs/hep-th/0010274}{{\ttfamily hep-th/0010274}}].

\bibitem{Pestun:2007rz}
V.~Pestun, \emph{{Localization of gauge theory on a four-sphere and
  supersymmetric Wilson loops}},
  \href{https://doi.org/10.1007/s00220-012-1485-0}{\emph{Commun. Math. Phys.}
  {\bfseries 313} (2012) 71} [\href{https://arxiv.org/abs/0712.2824}{{\ttfamily
  0712.2824}}].

\bibitem{Makeenko:2006ds}
Y.~Makeenko, P.~Olesen and G.~W. Semenoff, \emph{{Cusped SYM Wilson loop at two
  loops and beyond}},
  \href{https://doi.org/10.1016/j.nuclphysb.2006.05.002}{\emph{Nucl. Phys.}
  {\bfseries B748} (2006) 170}
  [\href{https://arxiv.org/abs/hep-th/0602100}{{\ttfamily hep-th/0602100}}].

\bibitem{Drukker:2011za}
N.~Drukker and V.~Forini, \emph{{Generalized quark-antiquark potential at weak
  and strong coupling}},
  \href{https://doi.org/10.1007/JHEP06(2011)131}{\emph{JHEP} {\bfseries 06}
  (2011) 131} [\href{https://arxiv.org/abs/1105.5144}{{\ttfamily 1105.5144}}].

\bibitem{Correa:2012nk}
D.~Correa, J.~Henn, J.~Maldacena and A.~Sever, \emph{{The cusp anomalous
  dimension at three loops and beyond}},
  \href{https://doi.org/10.1007/JHEP05(2012)098}{\emph{JHEP} {\bfseries 05}
  (2012) 098} [\href{https://arxiv.org/abs/1203.1019}{{\ttfamily 1203.1019}}].

\bibitem{Henn:2013wfa}
J.~M. Henn and T.~Huber, \emph{{The four-loop cusp anomalous dimension in
  $\mathcal{N} =$ 4 super Yang-Mills and analytic integration techniques for
  Wilson line integrals}},
  \href{https://doi.org/10.1007/JHEP09(2013)147}{\emph{JHEP} {\bfseries 09}
  (2013) 147} [\href{https://arxiv.org/abs/1304.6418}{{\ttfamily 1304.6418}}].

\bibitem{Correa:2012hh}
D.~Correa, J.~Maldacena and A.~Sever, \emph{{The quark anti-quark potential and
  the cusp anomalous dimension from a TBA equation}},
  \href{https://doi.org/10.1007/JHEP08(2012)134}{\emph{JHEP} {\bfseries 08}
  (2012) 134} [\href{https://arxiv.org/abs/1203.1913}{{\ttfamily 1203.1913}}].

\bibitem{Drukker:2012de}
N.~Drukker, \emph{{Integrable Wilson loops}},
  \href{https://doi.org/10.1007/JHEP10(2013)135}{\emph{JHEP} {\bfseries 10}
  (2013) 135} [\href{https://arxiv.org/abs/1203.1617}{{\ttfamily 1203.1617}}].

\bibitem{Correa:2012at}
D.~Correa, J.~Henn, J.~Maldacena and A.~Sever, \emph{{An exact formula for the
  radiation of a moving quark in N=4 super Yang Mills}},
  \href{https://doi.org/10.1007/JHEP06(2012)048}{\emph{JHEP} {\bfseries 06}
  (2012) 048} [\href{https://arxiv.org/abs/1202.4455}{{\ttfamily 1202.4455}}].

\bibitem{Gromov:2012eu}
N.~Gromov and A.~Sever, \emph{{Analytic Solution of Bremsstrahlung TBA}},
  \href{https://doi.org/10.1007/JHEP11(2012)075}{\emph{JHEP} {\bfseries 11}
  (2012) 075} [\href{https://arxiv.org/abs/1207.5489}{{\ttfamily 1207.5489}}].

\bibitem{Gromov:2013qga}
N.~Gromov, F.~Levkovich-Maslyuk and G.~Sizov, \emph{{Analytic Solution of
  Bremsstrahlung TBA II: Turning on the Sphere Angle}},
  \href{https://doi.org/10.1007/JHEP10(2013)036}{\emph{JHEP} {\bfseries 10}
  (2013) 036} [\href{https://arxiv.org/abs/1305.1944}{{\ttfamily 1305.1944}}].

\bibitem{Gromov:2015dfa}
N.~Gromov and F.~Levkovich-Maslyuk, \emph{{Quantum Spectral Curve for a cusped
  Wilson line in $ \mathcal{N}=4 $ SYM}},
  \href{https://doi.org/10.1007/JHEP04(2016)134}{\emph{JHEP} {\bfseries 04}
  (2016) 134} [\href{https://arxiv.org/abs/1510.02098}{{\ttfamily
  1510.02098}}].

\bibitem{Drukker:2006xg}
N.~Drukker and S.~Kawamoto, \emph{{Small deformations of supersymmetric Wilson
  loops and open spin-chains}},
  \href{https://doi.org/10.1088/1126-6708/2006/07/024}{\emph{JHEP} {\bfseries
  07} (2006) 024} [\href{https://arxiv.org/abs/hep-th/0604124}{{\ttfamily
  hep-th/0604124}}].

\bibitem{Giombi:2017cqn}
S.~Giombi, R.~Roiban and A.~A. Tseytlin, \emph{{Half-BPS Wilson loop and
  AdS$_2$/CFT$_1$}},
  \href{https://doi.org/10.1016/j.nuclphysb.2017.07.004}{\emph{Nucl. Phys.}
  {\bfseries B922} (2017) 499}
  [\href{https://arxiv.org/abs/1706.00756}{{\ttfamily 1706.00756}}].

\bibitem{Mazac:2018mdx}
D.~Mazac and M.~F. Paulos, \emph{{The analytic functional bootstrap. Part I: 1D
  CFTs and 2D S-matrices}},
  \href{https://doi.org/10.1007/JHEP02(2019)162}{\emph{JHEP} {\bfseries 02}
  (2019) 162} [\href{https://arxiv.org/abs/1803.10233}{{\ttfamily
  1803.10233}}].

\bibitem{Mazac:2018ycv}
D.~Mazac and M.~F. Paulos, \emph{{The analytic functional bootstrap. Part II.
  Natural bases for the crossing equation}},
  \href{https://doi.org/10.1007/JHEP02(2019)163}{\emph{JHEP} {\bfseries 02}
  (2019) 163} [\href{https://arxiv.org/abs/1811.10646}{{\ttfamily
  1811.10646}}].

\bibitem{Dolan:2011dv}
F.~A. Dolan and H.~Osborn, \emph{{Conformal Partial Waves: Further Mathematical
  Results}},  \href{https://arxiv.org/abs/1108.6194}{{\ttfamily 1108.6194}}.

\bibitem{Mazac:2016qev}
D.~Mazac, \emph{{Analytic bounds and emergence of AdS$_{2}$ physics from the
  conformal bootstrap}},
  \href{https://doi.org/10.1007/JHEP04(2017)146}{\emph{JHEP} {\bfseries 04}
  (2017) 146} [\href{https://arxiv.org/abs/1611.10060}{{\ttfamily
  1611.10060}}].

\bibitem{Beccaria:2017rbe}
M.~Beccaria, S.~Giombi and A.~Tseytlin, \emph{{Non-supersymmetric Wilson loop
  in $ \mathcal{N} $ = 4 SYM and defect 1d CFT}},
  \href{https://doi.org/10.1007/JHEP03(2018)131}{\emph{JHEP} {\bfseries 03}
  (2018) 131} [\href{https://arxiv.org/abs/1712.06874}{{\ttfamily
  1712.06874}}].

\bibitem{Kim:2017sju}
M.~Kim, N.~Kiryu, S.~Komatsu and T.~Nishimura, \emph{{Structure Constants of
  Defect Changing Operators on the 1/2 BPS Wilson Loop}},
  \href{https://doi.org/10.1007/JHEP12(2017)055}{\emph{JHEP} {\bfseries 12}
  (2017) 055} [\href{https://arxiv.org/abs/1710.07325}{{\ttfamily
  1710.07325}}].

\bibitem{Cooke:2017qgm}
M.~Cooke, A.~Dekel and N.~Drukker, \emph{{The Wilson loop CFT: Insertion
  dimensions and structure constants from wavy lines}},
  \href{https://doi.org/10.1088/1751-8121/aa7db4}{\emph{J. Phys.} {\bfseries
  A50} (2017) 335401} [\href{https://arxiv.org/abs/1703.03812}{{\ttfamily
  1703.03812}}].

\bibitem{Cavaglia:2018lxi}
A.~Cavagli{\`a}, N.~Gromov and F.~Levkovich-Maslyuk, \emph{{Quantum spectral
  curve and structure constants in $ \mathcal{N}=4 $ SYM: cusps in the ladder
  limit}}, \href{https://doi.org/10.1007/JHEP10(2018)060}{\emph{JHEP}
  {\bfseries 10} (2018) 060}
  [\href{https://arxiv.org/abs/1802.04237}{{\ttfamily 1802.04237}}].

\bibitem{Alday:2007he}
L.~F. Alday and J.~Maldacena, \emph{{Comments on gluon scattering amplitudes
  via AdS/CFT}},
  \href{https://doi.org/10.1088/1126-6708/2007/11/068}{\emph{JHEP} {\bfseries
  11} (2007) 068} [\href{https://arxiv.org/abs/0710.1060}{{\ttfamily
  0710.1060}}].

\bibitem{Bruser:2018jnc}
R.~Br{\"u}ser, S.~Caron-Huot and J.~M. Henn, \emph{{Subleading Regge limit from
  a soft anomalous dimension}},
  \href{https://doi.org/10.1007/JHEP04(2018)047}{\emph{JHEP} {\bfseries 04}
  (2018) 047} [\href{https://arxiv.org/abs/1802.02524}{{\ttfamily
  1802.02524}}].

\bibitem{Cavaglia:2020hdb}
A.~Cavagli{\`a}, D.~Grabner, N.~Gromov and A.~Sever, \emph{{Colour-Twist
  Operators I: Spectrum and Wave Functions}},
  \href{https://arxiv.org/abs/2001.07259}{{\ttfamily 2001.07259}}.

\bibitem{Dorn:2020meb}
H.~Dorn, \emph{{On anomalous conformal Ward identities for Wilson loops on
  polygon-like contours with circular edges}},
  \href{https://arxiv.org/abs/2001.03391}{{\ttfamily 2001.03391}}.

\bibitem{Liendo:2018ukf}
P.~Liendo, C.~Meneghelli and V.~Mitev, \emph{{Bootstrapping the half-BPS line
  defect}}, \href{https://doi.org/10.1007/JHEP10(2018)077}{\emph{JHEP}
  {\bfseries 10} (2018) 077}
  [\href{https://arxiv.org/abs/1806.01862}{{\ttfamily 1806.01862}}].

\bibitem{Liendo:2016ymz}
P.~Liendo and C.~Meneghelli, \emph{{Bootstrap equations for $ \mathcal{N} $ = 4
  SYM with defects}},
  \href{https://doi.org/10.1007/JHEP01(2017)122}{\emph{JHEP} {\bfseries 01}
  (2017) 122} [\href{https://arxiv.org/abs/1608.05126}{{\ttfamily
  1608.05126}}].

\bibitem{Gromov:2015vua}
N.~Gromov, F.~Levkovich-Maslyuk and G.~Sizov, \emph{{Pomeron Eigenvalue at
  Three Loops in $\mathcal N=$ 4 Supersymmetric Yang-Mills Theory}},
  \href{https://doi.org/10.1103/PhysRevLett.115.251601}{\emph{Phys. Rev. Lett.}
  {\bfseries 115} (2015) 251601}
  [\href{https://arxiv.org/abs/1507.04010}{{\ttfamily 1507.04010}}].

\bibitem{Alfimov:2014bwa}
M.~Alfimov, N.~Gromov and V.~Kazakov, \emph{{QCD Pomeron from AdS/CFT Quantum
  Spectral Curve}}, \href{https://doi.org/10.1007/JHEP07(2015)164}{\emph{JHEP}
  {\bfseries 07} (2015) 164} [\href{https://arxiv.org/abs/1408.2530}{{\ttfamily
  1408.2530}}].

\bibitem{Erickson:1999qv}
J.~K. Erickson, G.~W. Semenoff, R.~J. Szabo and K.~Zarembo, \emph{{Static
  potential in N=4 supersymmetric Yang-Mills theory}},
  \href{https://doi.org/10.1103/PhysRevD.61.105006}{\emph{Phys. Rev.}
  {\bfseries D61} (2000) 105006}
  [\href{https://arxiv.org/abs/hep-th/9911088}{{\ttfamily hep-th/9911088}}].

\bibitem{Leurent:2013mr}
S.~Leurent and D.~Volin, \emph{{Multiple zeta functions and double wrapping in
  planar $N=4$ SYM}},
  \href{https://doi.org/10.1016/j.nuclphysb.2013.07.020}{\emph{Nucl. Phys.}
  {\bfseries B875} (2013) 757}
  [\href{https://arxiv.org/abs/1302.1135}{{\ttfamily 1302.1135}}].

\bibitem{Marboe:2014gma}
C.~Marboe and D.~Volin, \emph{{Quantum spectral curve as a tool for a
  perturbative quantum field theory}},
  \href{https://doi.org/10.1016/j.nuclphysb.2015.08.021}{\emph{Nucl. Phys.}
  {\bfseries B899} (2015) 810}
  [\href{https://arxiv.org/abs/1411.4758}{{\ttfamily 1411.4758}}].

\bibitem{McGovern:2019sdd}
J.~McGovern, \emph{{Scalar Insertions in Cusped Wilson Loops in the Ladders
  Limit of Planar $N$=4 SYM}},
  \href{https://arxiv.org/abs/1912.00499}{{\ttfamily 1912.00499}}.

\bibitem{Giombi:2018qox}
S.~Giombi and S.~Komatsu, \emph{{Exact Correlators on the Wilson Loop in
  $\mathcal{N}=4$ SYM: Localization, Defect CFT, and Integrability}},
  \href{https://doi.org/10.1007/JHEP11(2018)123,
  10.1007/JHEP05(2018)109}{\emph{JHEP} {\bfseries 05} (2018) 109}
  [\href{https://arxiv.org/abs/1802.05201}{{\ttfamily 1802.05201}}].

\bibitem{Giombi:2018hsx}
S.~Giombi and S.~Komatsu, \emph{{More Exact Results in the Wilson Loop Defect
  CFT: Bulk-Defect OPE, Nonplanar Corrections and Quantum Spectral Curve}},
  \href{https://doi.org/10.1088/1751-8121/ab046c}{\emph{J. Phys.} {\bfseries
  A52} (2019) 125401} [\href{https://arxiv.org/abs/1811.02369}{{\ttfamily
  1811.02369}}].

\bibitem{Derkachov:2019tzo}
S.~Derkachov and E.~Olivucci, \emph{{Exactly solvable magnet of conformal spins
  in four dimensions}},  \href{https://arxiv.org/abs/1912.07588}{{\ttfamily
  1912.07588}}.

\bibitem{colortwist2}
A.~Cavagli{\`a}, N.~Gromov, F.~Levkovich-Maslyuk and A.~Sever,
  ``{\textit{Colour-Twist Operators II: Correlation Functions}}, to appear.''

\bibitem{Correa:2018fgz}
D.~Correa, M.~Leoni and S.~Luque, \emph{{Spin chain integrability in
  non-supersymmetric Wilson loops}},
  \href{https://doi.org/10.1007/JHEP12(2018)050}{\emph{JHEP} {\bfseries 12}
  (2018) 050} [\href{https://arxiv.org/abs/1810.04643}{{\ttfamily
  1810.04643}}].

\end{thebibliography}\endgroup

\end{document}